\documentclass[sigconf, nonacm,urlbreakonhyphens=false]{acmart}
\pdfoutput=1
\usepackage{subcaption}
\usepackage{pgfplots}
\usepackage{pgfplots}
\pgfplotsset{compat=newest}

\definecolor{colorClasses2}{HTML}{FF0000}
\definecolor{colorClasses8}{HTML}{0000FF}
\definecolor{colorClasses32}{HTML}{000000}

\pgfplotscreateplotcyclelist{myColorList}{%
  red,mark=triangle\\
  blue,mark=pentagon\\%
  black,mark=square\\%
  brown,mark=o\\%
  teal,mark=diamond\\%
  violet,mark=Mercedes star\\%
  orange,mark=star\\%
  magenta,mark=|\\%
}

\pgfplotsset{
  mark repeat*/.style={
    scatter,
    scatter src=x,
    scatter/@pre marker code/.code={
      \pgfmathtruncatemacro\usemark{
        or(mod(\coordindex,#1)==0, (\coordindex==(\numcoords-1))
      }
      \ifnum\usemark=0
        \pgfplotsset{mark=none}
      \fi
    },
    scatter/@post marker code/.code={}
  },
  log x ticks with fixed point/.style={
      xticklabel={
        \pgfkeys{/pgf/fpu=true}
        \pgfmathparse{exp(\tick)}%
        \pgfmathprintnumber[fixed relative, precision=3]{\pgfmathresult}
        \pgfkeys{/pgf/fpu=false}
      }
  },
  log y ticks with fixed point/.style={
      yticklabel={
        \pgfkeys{/pgf/fpu=true}
        \pgfmathparse{exp(\tick)}%
        \pgfmathprintnumber[fixed relative, precision=3]{\pgfmathresult}
        \pgfkeys{/pgf/fpu=false}
      }
  },
  every axis/.style={scale only axis},
  major grid style={thin,dotted},
  minor grid style={thin,dotted},
  ymajorgrids,
  yminorgrids,
  every axis/.append style={
    line width=0.9pt,
    tick style={
      line cap=round,
      thin,
      major tick length=4pt,
      minor tick length=2pt,
    },
    mark options={solid},
  },
  legend cell align=left,
  legend style={
    line width=0.7pt,
    /tikz/every even column/.append style={column sep=2mm,black},
    /tikz/every odd column/.append style={black},
    mark options={solid},
    font=\footnotesize,
  },  
  dominanceMap/.style={
      patch,
      patch type=rectangle,
      shader=flat corner,
  },
  title style={yshift=-2pt},
  enlarge x limits=0.04,
  scale only axis,
  /pgf/number format/1000 sep={},
  axis lines*=left,
  xlabel near ticks,
  ylabel near ticks,
  axis lines*=left,
  label style={font=\footnotesize},
  every axis y label/.append style={yshift=-1pt,inner sep=0,outer sep=0},
  tick label style={font=\footnotesize},
  cycle list name=myColorList,
}

\usepackage{dsfont}
\usepackage{amsfonts}
\usepackage{xcolor}
\usepackage{booktabs}
\usepackage[linesnumbered,ruled,vlined]{algorithm2e}
\DontPrintSemicolon

\SetCommentSty{mycommfont}
\usepackage{multirow, makecell}
\usepackage{mathtools}
\usepackage{hyperref}
\usepackage[switch]{lineno}
\usepackage[capitalise,noabbrev]{cleveref}
\DeclarePairedDelimiter{\ceil}{\lceil}{\rceil}

\DeclareMathOperator*{\argmin}{arg\,min}
\def\lmax{ℓ_{\max}}

\newcommand{\mytitle}{Learned Static Function Data Structures}

\newcommand{\Oh}[1]{\mathcal{O}\!\left( #1\right)}

\newcommand{\changed}[1]{#1}

\usepackage{newunicodechar}
\usepackage{silence}
\newunicodechar{♢}{\tikz \node[inner sep=1.5,draw,diamond] {};}
\newunicodechar{☆}{\tikz \node[inner sep=1,draw,star,star point ratio=2] {};}
\newunicodechar{△}{\triangle}
\newunicodechar{⬜}{\kern 0.5pt\tikz \node[inner sep=1.7,draw,regular polygon,regular polygon sides=4] {};\kern 0.5pt}
\newunicodechar{◯}{\tikz[baseline=-3pt] \node[inner sep=1.7,draw,cloud,cloud puffs=4,cloud puff arc=190] {};}

\newunicodechar{⊥}{\bot}
\newunicodechar{•}{\item}
\newunicodechar{✓}{\checkmark}
\newcommand{\xmark}{\ding{55}}%
\newunicodechar{✗}{\xmark}
\newunicodechar{…}{\dots}
\newunicodechar{≔}{\coloneqq}
\newunicodechar{⁻}{^-}
\newunicodechar{⁺}{^+}
\newunicodechar{₋}{_-}
\newunicodechar{₊}{_+}
\newunicodechar{ℓ}{\ell}
\newunicodechar{•}{\item}
\newunicodechar{…}{\dots}
\newunicodechar{≔}{\coloneqq}
\newunicodechar{≤}{\leq}
\newunicodechar{≥}{\geq}
\newunicodechar{≰}{\nleq}
\newunicodechar{≱}{\ngeq}
\newunicodechar{⊕}{\oplus}
\newunicodechar{⊗}{\otimes}
\newunicodechar{≠}{\neq}
\newunicodechar{¬}{\neg}
\newunicodechar{≡}{\equiv}
\newunicodechar{₀}{_0}
\newunicodechar{₁}{_1}
\newunicodechar{₂}{_2}
\newunicodechar{₃}{_3}
\newunicodechar{₄}{_4}
\newunicodechar{₅}{_5}
\newunicodechar{₆}{_6}
\newunicodechar{₇}{_7}
\newunicodechar{₈}{_8}
\newunicodechar{₉}{_9}
\newunicodechar{ₚ}{_p}
\newunicodechar{ₙ}{_n}
\newunicodechar{ₐ}{_a}
\newunicodechar{ₑ}{_e}
\newunicodechar{ₕ}{_h}
\newunicodechar{ₖ}{_k}
\newunicodechar{ₗ}{_l}
\newunicodechar{ₘ}{_m}
\newunicodechar{ₛ}{_s}
\newunicodechar{ₜ}{_t}
\newunicodechar{ₓ}{_x}
\newunicodechar{⁰}{^0}
\newunicodechar{¹}{^1}
\newunicodechar{²}{^2}
\newunicodechar{³}{^3}
\newunicodechar{⁴}{^4}
\newunicodechar{⁵}{^5}
\newunicodechar{⁶}{^6}
\newunicodechar{⁷}{^7}
\newunicodechar{⁸}{^8}
\newunicodechar{⁹}{^9}
\newunicodechar{ⁿ}{^n}
\newunicodechar{∈}{\in}
\newunicodechar{∉}{\notin}
\newunicodechar{⊂}{\subset}
\newunicodechar{⊃}{\supset}
\newunicodechar{⊆}{\subseteq}
\newunicodechar{⊇}{\supseteq}
\newunicodechar{⊄}{\nsubset}
\newunicodechar{⊅}{\nsupset}
\newunicodechar{⊈}{\nsubseteq}
\newunicodechar{⊉}{\nsupseteq}
\newunicodechar{∪}{\cup}
\newunicodechar{∩}{\cap}
\newunicodechar{∀}{\forall}
\newunicodechar{∃}{\exists}
\newunicodechar{∄}{\nexists}
\newunicodechar{∨}{\vee}
\newunicodechar{∧}{\wedge}
\newunicodechar{⊼}{\bar{\wedge}}
\newunicodechar{⊽}{\bar{\vee}}
\newunicodechar{∧}{\wedge}
\newunicodechar{ℝ}{\mathbb{R}}
\newunicodechar{ℕ}{\mathbb{N}}
\newunicodechar{𝔼}{\mathbb{E}}
\newunicodechar{𝔽}{\mathbb{F}}
\newunicodechar{ℚ}{\mathbb{Q}}
\newunicodechar{ℤ}{\mathbb{Z}}
\newunicodechar{ℂ}{\mathbb{C}}
\newunicodechar{𝒪}{\mathcal{O}}
\newunicodechar{⌊}{\lfloor}
\newunicodechar{⌋}{\rfloor}
\newunicodechar{⌈}{\lceil}
\newunicodechar{⌉}{\rceil}
\newunicodechar{·}{\cdot}
\newunicodechar{∘}{\circ}
\newunicodechar{×}{\times}
\newunicodechar{↑}{\uparrow}
\newunicodechar{↓}{\downarrow}
\newunicodechar{→}{\rightarrow}
\newunicodechar{←}{\leftarrow}
\newunicodechar{⇒}{\Rightarrow}
\newunicodechar{⇐}{\Leftarrow}
\newunicodechar{↔}{\leftrightarrow}
\newunicodechar{⇔}{\Leftrightarrow}
\newunicodechar{↦}{\mapsto}
\newunicodechar{∅}{\varnothing}
\newunicodechar{∞}{\infty}
\newunicodechar{≅}{\cong}
\newunicodechar{≈}{\approx}
\newunicodechar{ℓ}{\ell}
\newunicodechar{𝟙}{\mathds{1}}
\newunicodechar{𝟘}{\mathds{0}}
\newunicodechar{↪}{\hookrightarrow}

\newunicodechar{α}{\alpha}
\newunicodechar{β}{\beta}
\newunicodechar{γ}{\gamma}
\newunicodechar{Γ}{\Gamma}
\newunicodechar{δ}{\delta}
\newunicodechar{Δ}{\Delta}
\newunicodechar{ε}{\varepsilon}
\newunicodechar{ζ}{\zeta}
\newunicodechar{η}{\eta}
\newunicodechar{θ}{\theta}
\newunicodechar{Θ}{\Theta}
\newunicodechar{ι}{\iota}
\newunicodechar{κ}{\kappa}
\newunicodechar{λ}{\lambda}
\newunicodechar{Λ}{\Lambda}
\newunicodechar{μ}{\mu}
\newunicodechar{ν}{\nu}
\newunicodechar{ξ}{\xi}
\newunicodechar{Ξ}{\Xi}
\newunicodechar{π}{\pi}
\newunicodechar{Π}{\Pi}
\newunicodechar{ρ}{\rho}
\newunicodechar{σ}{\sigma}
\newunicodechar{Σ}{\Sigma}
\newunicodechar{τ}{\tau}
\newunicodechar{υ}{\upsilon}
\newunicodechar{ϒ}{\Upsilon}
\newunicodechar{φ}{\phi}
\newunicodechar{ϕ}{\varphi}
\newunicodechar{Φ}{\Phi}
\newunicodechar{χ}{\chi}
\newunicodechar{ψ}{\psi}
\newunicodechar{Ψ}{\Psi}
\newunicodechar{ω}{\omega}
\newunicodechar{Ω}{\Omega}

\newunicodechar{𝒜}{\mathcal{A}}
\newunicodechar{ℬ}{\mathcal{B}}
\newunicodechar{𝒞}{\mathcal{C}}
\newunicodechar{𝒟}{\mathcal{D}}
\newunicodechar{ℰ}{\mathcal{E}}
\newunicodechar{ℱ}{\mathcal{F}}
\newunicodechar{𝒢}{\mathcal{G}}
\newunicodechar{ℋ}{\mathcal{H}}
\newunicodechar{ℐ}{\mathcal{I}}
\newunicodechar{𝒥}{\mathcal{J}}
\newunicodechar{𝒦}{\mathcal{K}}
\newunicodechar{ℒ}{\mathcal{L}}
\newunicodechar{ℳ}{\mathcal{M}}
\newunicodechar{𝒩}{\mathcal{N}}
\newunicodechar{𝒪}{\mathcal{O}}
\newunicodechar{𝒫}{\mathcal{P}}
\newunicodechar{𝒬}{\mathcal{Q}}
\newunicodechar{ℛ}{\mathcal{R}}
\newunicodechar{𝒮}{\mathcal{S}}
\newunicodechar{𝒯}{\mathcal{T}}
\newunicodechar{𝒰}{\mathcal{U}}
\newunicodechar{𝒱}{\mathcal{V}}
\newunicodechar{𝒲}{\mathcal{W}}
\newunicodechar{𝒳}{\mathcal{X}}
\newunicodechar{𝒴}{\mathcal{Y}}
\newunicodechar{𝒵}{\mathcal{Z}}
\newunicodechar{𝒶}{\mathcal{a}}
\newunicodechar{𝒷}{\mathcal{b}}
\newunicodechar{𝒸}{\mathcal{c}}
\newunicodechar{𝒹}{\mathcal{d}}
\newunicodechar{ℯ}{\mathcal{e}}
\newunicodechar{𝒻}{\mathcal{f}}
\newunicodechar{ℊ}{\mathcal{g}}
\newunicodechar{𝒽}{\mathcal{h}}
\newunicodechar{𝒾}{\mathcal{i}}
\newunicodechar{𝒿}{\mathcal{j}}
\newunicodechar{𝓀}{\mathcal{k}}
\newunicodechar{𝓁}{\mathcal{l}}
\newunicodechar{𝓂}{\mathcal{m}}
\newunicodechar{𝓃}{\mathcal{n}}
\newunicodechar{ℴ}{\mathcal{o}}
\newunicodechar{𝓅}{\mathcal{p}}
\newunicodechar{𝓆}{\mathcal{q}}
\newunicodechar{𝓇}{\mathcal{r}}
\newunicodechar{𝓈}{\mathcal{s}}
\newunicodechar{𝓉}{\mathcal{t}}
\newunicodechar{𝓊}{\mathcal{u}}
\newunicodechar{𝓋}{\mathcal{v}}
\newunicodechar{𝓌}{\mathcal{w}}
\newunicodechar{𝓍}{\mathcal{x}}
\newunicodechar{𝓎}{\mathcal{y}}
\newunicodechar{𝓏}{\mathcal{z}}

\newcommand\vldbdoi{10.14778/3796195.3796205}
\newcommand\vldbpages{917-930}
\newcommand\vldbvolume{19}
\newcommand\vldbissue{5}
\newcommand\vldbyear{2026}
\newcommand\vldbauthors{\authors}
\newcommand\vldbtitle{\shorttitle} 
\newcommand\vldbavailabilityurl{https://github.com/gvinciguerra/LearnedStaticFunction}
\newcommand\vldbpagestyle{empty}

\def\quantover{0.1}

\def\quantoverint{149}

\def\quantinfover{2.9}

\def\quantinfoverint{68}

\def\infmaxrfast{65}\def\infmaxrfast{65}\def\infmaxrfast{65}\def\infmaxrfast{65}\def\infmaxrfast{65}\def\infmaxrfast{65}\def\infmaxrfast{65}\def\infmaxrfast{65}

\def\allquerygov{12.2}

\def\allqueryburr{29.2}

\def\huffshanspeedup{81}

\def\filteredfanospeedup{44}

\def\huffshanspace{1.03}

\def\fiterrootspace{0.18}

\def\ourcsfover{3.6}

\def\ourcsfburrover{1.8}

\def\ourcsfgov{7.1}

\def\vlburrover{1.2}

\def\storageover{10.1}

\def\storageovergausszero{133}

\def\storageoverurls{14.8}

\def\spacemaxgausszerourls{56}

\def\querymax{3.478}

\def\querymin{0.3}

\def\infmaxr{88}

\def\infminr{39}

\def\covertrainmus{215}

\def\maxspacerw{94}

\def\minspacerw{37}

\def\urlspace{94.9}

\def\urlspacefactor{19.5}

\def\gauszerospace{99.98}

\def\gauszerospacefactor{5448}

\def\gausthreespace{67.7}

\def\nidsspace{54.9}

\def\covertypespace{67.2}

\def\songsentropysigma{34.4}

\def\songsspaceperc{37.1}

\def\songsclasses{82}

\def\gaussclasses{8}

\def\consgaussgov{1.49}

\def\gausstrainmax{0.003}

\def\consallburr{1777}

\def\consallgov{131}

\def\urlsentropy{0.99}

\def\urlsentropy{0.99}

\def\urlsabspace{0.053}

\def\urlsoverentropy{18.7}

\def\maxeceperc{0.6}

\usepackage{multicol}

\begin{document}
\title{\mytitle}%

\newif\ifarxiv
\arxivtrue
\ifarxiv
\titlenote{Accepted for publication in Proceedings of the VLDB Endowment (PVLDB), Vol. 19, ISSN 2150-8097.}
\else
\fi

\author{Stefan Hermann}
\orcid{0000-0001-9183-2926}
\affiliation{%
  \institution{Karlsruhe Institute of Technology}
  \city{Karlsruhe}
  \country{Germany}
}
\email{hermann@kit.edu}

\author{Hans-Peter Lehmann}
\orcid{0000-0002-0474-1805}
\affiliation{%
  \institution{Karlsruhe Institute of Technology}
  \city{Karlsruhe}
  \country{Germany}
}
\email{hans-peter.lehmann@kit.edu}

\author{Giorgio Vinciguerra}
\orcid{0000-0003-0328-7791}
\affiliation{%
  \institution{University of Pisa}
  \city{Pisa}
  \country{Italy}
}
\email{giorgio.vinciguerra@unipi.it}

\author{Stefan Walzer}
\orcid{0000-0002-6477-0106}
\affiliation{%
  \institution{Karlsruhe Institute of Technology}
  \city{Karlsruhe}
  \country{Germany}
}
\email{stefan.walzer@kit.edu}

\begin{abstract}
    We consider the task of constructing a data structure for associating a static set of keys with values, while allowing arbitrary output values for queries involving keys outside the set.
    Compared to hash tables, these so-called \emph{static function data structures} do not need to store the key set and thus use significantly less memory.
    Several techniques are known, with \emph{compressed} static functions approaching the zero-order empirical entropy of the value sequence.
    In this paper, we introduce \emph{learned} static functions, which use machine learning to capture correlations between keys and values.
    For each key, a model predicts a probability distribution over the values, from which we derive a key-specific prefix code to compactly encode the true value. The resulting codeword is stored in a classic static function data structure.
    This design allows learned static functions to break the zero-order entropy barrier while still supporting point queries.
    Our experiments show substantial space savings: up to one order of magnitude on real data, and up to three orders of magnitude on synthetic data.
\end{abstract}

\maketitle

\ifarxiv
\pagestyle{plain}
\else
\pagestyle{\vldbpagestyle}
\begingroup\small\noindent\raggedright\textbf{PVLDB Reference Format:}\\
\vldbauthors. \vldbtitle. PVLDB, \vldbvolume(\vldbissue): \vldbpages, \vldbyear.\\
\href{https://doi.org/\vldbdoi}{doi:\vldbdoi}
\endgroup
\begingroup
\renewcommand\thefootnote{}\footnote{\noindent
This work is licensed under the Creative Commons BY-NC-ND 4.0 International License. Visit \url{https://creativecommons.org/licenses/by-nc-nd/4.0/} to view a copy of this license. For any use beyond those covered by this license, obtain permission by emailing \href{mailto:info@vldb.org}{info@vldb.org}. Copyright is held by the owner/author(s). Publication rights licensed to the VLDB Endowment. \\
\raggedright Proceedings of the VLDB Endowment, Vol. \vldbvolume, No. \vldbissue\ %
ISSN 2150-8097. \\
\href{https://doi.org/\vldbdoi}{doi:\vldbdoi} \\
}\addtocounter{footnote}{-1}\endgroup

\ifdefempty{\vldbavailabilityurl}{}{
\vspace{.3cm}
\begingroup\small\noindent\raggedright\textbf{PVLDB Artifact Availability:}\\
The source code, data, and/or other artifacts have been made available at \url{\vldbavailabilityurl}.
\endgroup
}
\fi

\section{Introduction}
\label{sec:introduction}

The standard way for storing a map $f: K → V$ from keys to values is a hash table \cite{koppl2022fast,bender2021all,Knuth1998art,pagh2004cuckoo}.
However, when $f$ is static and queries are restricted to keys in~$K$ (with arbitrary outputs allowed otherwise), specialised data structures called \emph{static functions} or \emph{retrieval data structures} can be used instead~\cite{DP:Succinct:2008,BKZ:A_Practical:2005,Porat09bloomreplacement,BPZ:Practical:2013,GenuzioOV20minimal,DBLP:journals/jea/GrafL22,DBLP:journals/jea/GrafL20,DillingerHSW2022burr,vigna2025sharding}.
Their main advantage is that the keys do not have to be stored, and the mapping can thus be represented in a little more than $|K|\log₂|V|$~bits.%
\footnote{How precisely an SF can achieve this using systems of linear equations is beside the point of this introduction. Some details are unpacked in \cref{s:comp-retr} as needed.}
Consider, for instance, a set $K$ of $n$ URLs annotated with labels from $V = \{\textsc{good},\textsc{ai-slop}, \textsc{phishing}\}$. A static function could represent this mapping using $n\log₂ 3 < 2n$ bits, regardless of the URL lengths.
\changed{In contrast, any encoding of the set of all key-value pairs (e.g.\ using a hash table) would need orders of magnitude more space.}

\begin{figure}
    \includegraphics[scale=0.85]{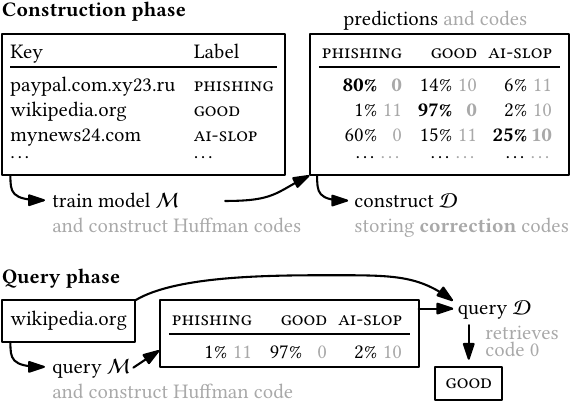}
    \caption{
        Architecture of learned static functions.
        \changed{In gray, a simple implementation that stores Huffman codes in the auxiliary data structure $\mathcal{D}$.}
    }
    \label{fig:architecture}
\end{figure}

For random functions with domain $K$ and range $V$, we cannot do better than $|K|\log₂|V|$~bits on average. However, when some values from $V$ appear much more frequently than others,  the space usage can be further reduced using \emph{compressed static functions} (CSFs) to approximately $nH_0$ bits, where $H_0=\sum_{v\in V}p_v\log_2 \left(1/p_v\right)$ denotes the zero-order empirical entropy of the value sequence, and $p_v ∈ [0,1]$ is the relative frequency of $v$~\cite{HreinssonKP09compressedfun,BelazzouguiV13compressedfun,GenuzioOV20minimal}.
In the example, if $95\%$ of the URLs are \textsc{good}, $4\%$ are \textsc{ai-slop} and $1\%$ are \textsc{phishing}, a CSF could  achieve as little as $0.32n$~bits.

If $f$ exhibits low $H₀$ but is otherwise random, we have again hit an information-theoretic barrier.
However, if $f$ stores real-world data, then knowing a key may allow us to make an informed guess about its associated value.
In this paper, we leverage such key-specific predictions to achieve significant compression, even when each value appears equally often overall, i.e., even when $H₀ ≈ \log₂|V|$.
Going back to our example, we might observe that most \texttt{.edu} URLs are \textsc{good}, while URLs containing sneaky misspellings are more likely to be \textsc{phishing}.
Such correlations are often subtle, domain-specific, and difficult to capture with handcrafted rules, making learning-based approaches a natural fit.
We hence propose \emph{learned static functions}~(LSFs), which involve two components, as illustrated in \cref{fig:architecture}.
\begin{itemize}
    • First, a model $ℳ$ that is specific to the use case and trained to predict $f(k)$ given $k ∈ K$. Technically, $ℳ$ must output a probability distribution on $V$ reflecting what it believes $f(k)$ to be. Additional data related to $k$ can inform the prediction, provided it is also available at query time.\footnote{This makes no conceptual difference as any such information can be considered to be part of the key. An example could be the key being a file name and queries reading the file to extract additional information.}
    • Second, an auxiliary data structure $\mathcal{D}$, much like a CSF, \changed{intuitively responsible for correcting all the mistakes made by $ℳ$.
    This suggests that if $ℳ$ is right about most keys, then~$\mathcal{D}$ needs only to store a small set of exceptions. Note, however, that technically the “rightness” of $ℳ$ about a given key is not categorical but a matter of degree (explained more precisely in \cref{sec:prelim}).}
\end{itemize}

\paragraph{When LSFs Might be Useful\@.}
\changed{Like all SFs, an LSF faithfully returns $f(k)$ whenever $k \in K$, but returns an arbitrary value for $k ∉ K$.}
Among contexts where this is acceptable, LSFs are particularly well-suited for scenarios where space efficiency is paramount, such as when the key set is so large that traditional data structures like CSFs no longer fit in fast, local, or cheap memory.
LSFs are especially effective when the mapping from keys to values exhibits structure that can be learned, in the sense that the combined encoding of the model $ℳ$ and the data structure $\mathcal{D}$ meaningfully undercuts $H_0$ that would be achievable with a CSF.

\paragraph{Applications\@.}
SFs and CSFs have applications in areas such as bioinformatics \cite{ShibuyaBK22kmer,ColemanRLLS2023caramel}, log management~\cite{reichinger2024copr}, systems for ML~\cite{ColemanRLLS2023caramel}, data structures for range queries \cite{alstrup2001optimal, mortensen2005dynamic}, perfect hashing~\cite{BPZ:Practical:2013,GenuzioOV20minimal,BelazzouguiBPV2011theoryPractice,FerraginaLSV2023lemon, hermann2025morphishash,DBLP:conf/alenex/Lehmann0W24}, prefix search \cite{belazzougui2010fast}, and approximate membership  \cite{DP:Succinct:2008,DBLP:journals/jea/GrafL20,DBLP:journals/jea/GrafL22,DillingerHSW2022burr} (Bloom filters), which themselves have widespread~use.

LSFs open new opportunities in contexts where keys and values exhibit correlations.
For example, we experiment with a dataset where URLs are labelled with $\{\textsc{phishing},\textsc{good}\}$.
Such a labelling might be generated by a web-crawler at an early stage to later also detect clusters of sites linked to phishing.
Both labels appear similarly often.
Our LSF represents the labelling using \urlsabspace~bits per key, while any CSF requires at least \urlsentropy~bits per key\,---\,a compression by a factor of~\urlsoverentropy.
Our experiments encompass additional real-world and synthetic datasets.
In the following, we outline possibilities which exceed those we experimented with.

\begin{itemize}
    \item \emph{Tabular Data.}
        In many structured datasets, such as relational databases, the value of one column can often be inferred from others.
        For example, in a product catalogue, the category of an item (e.g., books, clothing, electronics) can often be inferred from fields like the title or brand.
        LSFs offer a natural solution to compress and efficiently retrieve these predictable values.
        Previous work exploited such learnability in tabular data in a lossy way~\cite{DeepSqueeze,SPARTAN}, without point query support~\cite{SQUISH,DaviesM99}, or assuming linear correlation only~\cite{LiuSRK24}.
        \changed{Additional opportunities arise from exploiting correlations across tables. For example, customer attributes such as demographics can help predict order-related columns such as purchased product, payment method, or delivery time. When tables are joined and materialised (e.g., in views), LSF can help reduce storage space.}

    \item \emph{Hyphenation.}
        Imagine $f$ maps English words to their correct \hbox{hy-phen-ation}, i.e., a set of positions where a word break is allowed.
        It is not hard to imagine cases where an app could use a corresponding data structure but would only want to use negligible memory to store it and would not require large throughput. 
        Given that hyphenation sounds highly learnable, an LSF might be useful.
    \item \emph{Map Annotations.}
        Imagine a road network where some information is useful but unlikely to be requested frequently, such as whether some footpath is wheelchair accessible or whether some road is winter maintained.
        Geographic regularities might make such data highly learnable.
        We experiment with a related (though different) dataset regarding forest types.
\item \emph{Endgame Tablebases.}
        When analysing games with complete information such as chess, it can be useful to tabulate information for vast sets of game states (especially states with few remaining pieces), storing for each the best possible move or simply the outcome from $\{\textsc{draw},\textsc{win},\textsc{loss}\}$ assuming optimal play \cite{gomboc2024comparing}.
        Such labels are clearly learnable, and thus LSFs could help compress tablebases.
        It seems, however, that the LSF would also have to be made aware of complicated game-specific symmetries that are well-known but beyond the scope of this paper.
    \item \emph{$k$-mer Annotations.}~\looseness=-1
        Over the past 20 years, the cost of genome sequencing has dropped by roughly four orders of magnitude, leading to a growth of bioinformatics datasets that outpace the rate at which computational hardware becomes cheaper.
        We believe that LSFs can play a role in conserving memory, e.g., when storing annotations of large sets of DNA snippets, called $k$-mers.
        The annotations might reflect taxonomic origin (think “mouse or rat”), whether the snippet is coding or non-coding, its function (related GO~\cite{gene2021gene} term or KEGG~\cite{kanehisa2021kegg} pathway), or  whether it is unique in the reference genome.
        All of these are learnable in some contexts.
\end{itemize}

\paragraph{Our Contributions\@.}
We achieve the following:
\begin{itemize}
    \item We formally introduce the concept of learned static functions (LSFs) and give the first practical implementation.
    \item As an intermediate problem, we extend the applicability of BuRR~\cite{DillingerHSW2022burr}, a highly space-efficient practical SF, by adapting it to implement a variable-length static function (VL-SF) storing a map $f : K → \{0,1\}^*$ from keys to variable-length bit strings. 
    \item We address a key inefficiency in the design of   LSFs and existing CSFs~\cite{HreinssonKP09compressedfun,GenuzioOV20minimal} based on VL-SF by introducing a “generalised filter trick”, which saves up to $\frac 12$ bits per key.\footnote{The case where up to $\frac 12$ bits per key can be saved is the second example in \cref{fig:suboptimal-prefix-codes}.}
    \item We evaluate several machine learning (ML) models to minimise the space of our LSF, showing how model choice impacts performance and storage.
    \item We demonstrate the effectiveness of our overall approach on real-world datasets spanning diverse domains such as geographic information systems, network security, phishing detection, and music, as well as synthetic datasets. Our experiments show space savings of up to one order of magnitude on real data, and up to three orders of magnitude on synthetic data.
\end{itemize}

\paragraph{Outline\@.}
We start with preliminaries in \cref{sec:prelim}.
We then introduce our new type of data structures, learned static functions, in \cref{sec:lsfs}.
In \cref{s:comp-retr}, we build a VL-SF based on BuRR \cite{DillingerHSW2022burr}.
In \cref{s:lsf}, we explain the implementation of an LSF based on VL-SFs and look at engineering challenges.
We perform an evaluation using different ML models in \cref{sec:experiments}.
In \cref{sec:related}, we explain related work from the literature and compare it to our new data structure.
Finally, we conclude the paper in \cref{sec:conclusion}.

\section{Preliminaries}
\label{sec:prelim}
A \emph{static function} (SF), also called a \emph{static retrieval data structure}, is a data structure constructed for a function $f: K → V$ where $K$ is a set of $n$ \emph{keys} from a universe $U$ and $V$ is a (typically small) set of \emph{values} or \emph{labels}.
After construction, it cannot be modified but just queried with $k ∈ U$.
If $k ∈ K$, then the answer must be $f(k)$. If $k ∈ U \setminus K$, then the answer is arbitrary. In \cref{tab:types-of-SF} we summarise variants of SFs discussed below.

\def\vertbrace{$\left.\makebox(0pt,12pt){}\right\}$}
\begin{table}[tb]
    \caption{
        Types of static functions (SFs).
    }
	\label{tab:types-of-SF}
  \centering
    \begin{tabular}{rlll}
    	\toprule
    	Type & Context & Ideal space (bits) & References\\
    	\midrule
          SF
        & $f: K → V$
        & $n \log₂ |V|$
        & \multirow{2}{*}{\vertbrace\parbox{1.7cm}{\cite{DP:Succinct:2008,BKZ:A_Practical:2005,Porat09bloomreplacement,BPZ:Practical:2013,GenuzioOV20minimal,DBLP:journals/jea/GrafL22,DBLP:journals/jea/GrafL20,DillingerHSW2022burr}}}
        \\
        $r$-bit SF
        & $f: K → \{0,1\}^r$
        & $nr$\\
        VL-SF
        & $f:K→\{0,1\}^*$
        & $\sum_{k ∈ K} |f(k)|$
        & \multirow{2}{*}{\vertbrace\cite{GenuzioOV20minimal,HreinssonKP09compressedfun,BelazzouguiV13compressedfun,ColemanRLLS2023caramel}}\\
        CSF
        & $f: K → V$
        & $nH₀$ %
        & \\
    	\midrule
        \multirow{2}{*}{LSF}
        & $f: K → V,$
        & $S(f,ℳ)$
        & 
        \multirow{2}{*}{\textsc{new}}\\
        & $ℳ = (μ_k)_{k ∈ K}$
        & +~|enc($ℳ$)|
        &\\
    	\bottomrule
    \end{tabular}
\end{table}

\paragraph{Related: Filter Data Structures\@.}\label{par:related-data-structures}
In a sense, an SF must “forget” parts of the data for which it was constructed to be space-efficient.
It shares this property with Bloom filters \cite{bloom1970space} (filters for short, also known as approximate membership query data structures).
A filter represents a set $M ⊆ U$ and answers membership queries “$k ∈ M$?” with one-sided errors, i.e., any $k ∉ M$ may be erroneously reported to be present in $M$ with a small false-positive probability $ε$.
Any SF can be used as a filter \cite{DP:Succinct:2008,DBLP:journals/jea/GrafL22,DBLP:journals/jea/GrafL20,DillingerHSW2022burr}, while filters are helpful when constructing CSFs \cite{GenuzioOV20minimal,HreinssonKP09compressedfun,BelazzouguiV13compressedfun,ColemanRLLS2023caramel}.

\paragraph{$r$-bit Static Functions\@.}
Most uncompressed SFs assume $V = \{0,1\}^r$ for some $r ∈ ℕ$; we call them $r$-bit SFs.%
\footnote{An exception is an SF using $V = \{0,1,2\}$ \cite[Section 7.2]{GenuzioOV20minimal}.}
The case of $r = 1$, i.e.\ $V = \{0,1\}$, has also been called the relative membership problem~\cite{BelazzouguiV13compressedfun} and is closely related to Bloom filters with a false positive free zone \cite{DBLP:journals/tnsm/KissHTRR21}.
Succinct $r$-bit SFs using $(1+o(1))rn$ bits are known; we will build upon one such SF called \emph{BuRR} \cite{DillingerHSW2022burr} in \cref{s:comp-retr}.

\paragraph{Variable-Length Static Functions (VL-SF)\@.}
Given a space-efficient \mbox{$1$-bit} SF, it is not hard to construct an SF for the case where values are variable-length bit strings, i.e. $V = \{0,1\}^*$, while using close to $\sum_{k ∈ K} |f(k)|$~bits.
Conceptually, we treat a key~$k$ associated with an $l$-bit string $b₁…b_l$ as $l$ independent keys $(k,1),…,(k,l)$, where $(k,i)$ is associated with $b_i$. If we know a key $k$ and $l$, then we can recover $f(k)$ by internally querying $(k,1),…,(k,l)$. A challenge is to use this idea without losing a factor of $l$ in running time. We comment on these challenges in the case of BuRR in \cref{ss:comp-retr-vlr}.

Note that if $l$ is not known when querying $k$, we get a stream of bits that begins with $f(k)$ and continues with garbage bits. The full VL-SF needs to detect when the meaningful part of the stream ends. This is easy if we are storing prefix codes, but otherwise additional space might be required.

We remark that storing variable-length bit strings efficiently is non-trivial even if $K = \{1,…,n\}$, i.e., when storing an array of variable-length bit strings with random access, see e.g., \cite{DBLP:conf/dcc/Kulekci14,Lurpicz2023pachash} and \cite[Section 3.2]{Navarro2016book}. See also \cite{DBLP:conf/focs/Patrascu08} for compressed arrays.

\paragraph{Compressed Static Functions (CSFs)\@.}
It is easy to build CSFs from VL-SFs. Given $f: K → V$, we consider the relative frequencies $p_v := |f^{-1}(v)|/|K|$ with which $v$ occurs and construct a corresponding optimal prefix code for $V$ (e.g.\ a Huffman code \cite{huffman2007method}).
We then use a VL-SF to store the codeword of $f(k)$ for each $k ∈ K$.
The combined codeword length is then $|K|(H₀+δ)$ bits, where $δ ∈ [0,1]$ is a term accounting for the redundancy of the prefix code~\cite{Gallager78}.
The full CSF must also store the prefix code (e.g., as a Huffman tree), which takes negligible space.
GOV~\cite{GenuzioOV20minimal} is the current state-of-the-art practical CSF.
\footnote{CARAMEL~\cite{ColemanRLLS2023caramel} implements a CSF that supports the storage of tuples rather than single values. Since it builds directly on GOV’s implementation \changed{and, in our setting without tuples, would perform equivalently to GOV,} we do not discuss it further.\label{foot:caramel}}
A noteworthy corner case for classical CSFs is when $H₀$ is close to zero, which may happen if the same element of $V$ is assigned to most keys from $K$.
In that case, the CSF as described above still requires at least $1$ bit per key and thus is not space efficient.
The problem can be fixed by using a Bloom filter that represents the small set of keys not mapping to the majority element \cite{HreinssonKP09compressedfun,GenuzioOV20minimal,BelazzouguiV13compressedfun,ColemanRLLS2023caramel, ShibuyaBK22kmer}.
We explain and expand upon this \emph{filter trick} in \cref{s:lsf}.

\changed{
\paragraph{Surprisal.}\label{par:surprisal}
Given a probability distribution $μ$ and an outcome $x$ with probability $p = μ(x)$, then $\log₂(1/p)$ is called the \emph{surprisal} (or \emph{information content}) of $x$ given $μ$.
Note that surprisal can arise even when $x$ is not random:
If Alice, when asked about the capital of Canada, assigns a credence of $75\%$ to Toronto and $25\%$ to Ottawa, then her surprisal when learning the truth (Ottawa) is $\log₂(4) = 2$ bits. A different observer may experience different surprisal.
}

\section{Learned Static Functions}\label{sec:lsfs}

We are now ready to introduce the concept of learned static functions (LSFs). An LSF represents a mapping $f: K → V$ in a context where we can construct an ML model~$ℳ$ that, given any $k ∈ K$, outputs a probability distribution $μ_k$ on~$V$, intuitively reflecting what $ℳ$ believes $f(k)$ to be.%
\footnote{We could also work with a weaker model that merely outputs a single value $v_k ∈ V$ and a confidence $p ∈ (0,1)$ that $v_k = f(k)$.
To obtain a distribution $μ_k$ on all of $V$, we could simply allocate the remaining probability mass $1-p$ to $V \setminus \{v_k\}$ according to the relative frequencies or in some other ad-hoc way. However, this is undoubtedly less powerful in some contexts.}
From each distribution~$μ_k$, we derive a prefix code and store the codeword corresponding to $f(k)$ in an auxiliary data structure~$\mathcal{D}$.
This step is analogous in spirit to a CSF but with a crucial difference: the prefix code is determined by $μ_k$ and is specific to each key, while a CSF uses the same prefix code for every key.
Importantly, the prefix code need not be stored: a query for $k$ will consult $ℳ$ again, obtain the same probability distribution $μ_k$ and compute the same prefix code, which allows decoding $f(k)$ from the bits stored in~$\mathcal{D}$.

Ideally, the storage cost for each key $k$ is $\log₂(1/μ_k(f(k)))$~bits, which is the surprisal of the outcome $f(k)$ under the probability distribution $μ_k$.\footnote{This is sometimes called the categorical cross entropy, i.e.\ the cross entropy of $μ_k$ relative to the distribution that assigns probability $1$ to $f(k)$.}
Summing over all keys yields the total space cost
\begin{equation}
    S(ℳ,f) := \sum_{k ∈ K} \log₂(1/μ_k(f(k))).
	\label{eq:loss-function}
\end{equation}

While this suggests that $S(ℳ,f)$ can serve as a loss function when training $ℳ$, minimising it alone does not fully capture the actual optimisation goal in the LSF setting, which  differs from classical supervised learning in two important ways. 
First, the model does not have to generalise well beyond $K$, i.e., overfitting to $K$ is not a problem but actually desirable to some extent.
Second, the size $|\textrm{enc}(ℳ)|$ of the model $ℳ$ itself contributes to the total space budget.
The actual optimisation goal is therefore minimising
\begin{equation}
    \text{space} ≈ \Sigma := S(ℳ,f) + |\textrm{enc}(ℳ)|.
    \label{eq:space}
\end{equation}
Therefore, for small- and medium-sized datasets, the most effective choice is typically a simple model or heuristic rather than a deep neural network, even if the latter yields a smaller value for $S(ℳ,f)$.
That said, in scenarios where a suitable model is already available (e.g., trained for a related upstream task), it can be reused for LSF at no additional storage cost, making the use of complex models more attractive in such cases.

Defining LSFs conceptually is only the first step.
To make LSFs practical, we must now design the auxiliary data structure $\mathcal{D}$.
To this end, we develop the following:
\begin{itemize}
  \item A space-efficient VL-SF called VL-BuRR (\Cref{s:comp-retr}).
  \item Based on VL-BuRR, a \emph{weighted filter data structure} where a weight $r(k)$ controls the false-positive probability $2^{-r(k)}$, individually for each key $k ∈ U$. (\Cref{ss:weighted-filters})
  \item Based on the weighted filter, a \emph{weighted relative membership}~(WRM) data structure, or equivalently, a $1$-bit SF where each key~$k$ has a weight~$p(k)$ that indicates the probability that $k$ is assigned a value of $1$. If the weights are provided by an ML model, we have a $1$-bit LSF. (\Cref{ss:WRM})
  \item A general LSF that conceptually reconstructs a prefix code of a key's value by using an adaptive sequence of $1$-bit LSF queries. (\Cref{ss:LSF-construction})
\end{itemize}
An advantage of our method is that prefix codes are not stored directly, and any inefficiencies in the prefix code are not reflected in the final space consumption.\footnote{For example, with a confident model prediction such as $μ_k(f(k))=0.9$, the ideal cost for key $k$ is only $\log_2(1/μ_k(f(k)))≈ 0.05$ bits, which we get close to, while a prefix code still requires at least 1 full bit.} By exploiting this flexibility:
\begin{itemize}
    \item We adopt prefix codes of suboptimal length that are much faster to construct than optimal prefix codes such as Huffman. (\Cref{ss:comp-retr-code})
\end{itemize}
In practice, the space required by $\mathcal{D}$ exceeds $S(ℳ,f)$ by at most \storageover \%, provided that $S(ℳ,f)$ is large enough to render lower-order overheads negligible.

\section{Space-Efficient VL-SF\MakeLowercase{s}}
\label{s:comp-retr}

In this section, we present a new design for VL-SFs, i.e. data structures that represent a mapping $f: K \rightarrow \{0,1\}^*$ where the values are variable-length bit strings.
Previous practical approaches incur a multiplicative space overhead of at least 2.4\% over the ideal cost of storing $\sum_{k \in K}|f(k)|$ bits \cite{GenuzioOV20minimal}.
Our approach builds upon BuRR~\cite{DillingerHSW2022burr} and extends it to efficiently handle variable-length values, achieving an overhead of at most \vlburrover\%.

We begin by briefly recalling BuRR in~\Cref{ss:burr}, and then we describe our new VL-SF construction in \Cref{ss:comp-retr-vlr}.

\begin{figure}
    \includegraphics[scale=0.9]{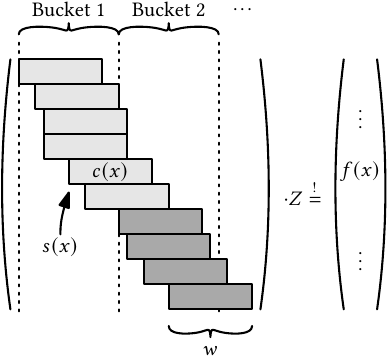}
    \caption{
        Structure of the BuRR equation system \cite{DillingerHSW2022burr}.
    }
    \label{fig:burr}
\end{figure}

\subsection{1-bit BuRR}
\label{ss:burr}

BuRR \cite{DillingerHSW2022burr} is an $r$-bit SF.
For our purposes, we recall its 1-bit variant, which stores a mapping $f: K \rightarrow \{0,1\}$ for a given set $K$ of $n$ keys.
The construction of BuRR is based on linear algebra over the field $\mathbb{F}_2 = \{0,1\}$.
The mapping itself is represented by a binary vector $Z \in \{0,1\}^m$, with $m\geq n$ rows.
To query a key~$k$, we first determine a hash row vector $h(k) \in \{0,1\}^{1 \times m}$, and then we compute the dot product $h(k)Z$ (modulo 2) to obtain $f(k)$.
To construct $Z$, we must ensure that $h(k)Z=f(k)$ for all keys~$k$.
This corresponds to solving the linear system $HZ=F$, where $H\in \{0,1\}^{n\times m}$ is the matrix with a row $h(k)$ for each $k\in K$, and $F\in\{0,1\}^n$ is the vector of all values.
Even if $m = n$, there is a chance of $≈ 0.289$ that the system is solvable if $h$ is fully random~\cite{cooper2000rank}.
We then obtain an SF that requires just $m$ bits to store $Z$.
However, the query time, i.e., the cost of computing $h(k)Z$, is linear in $m$.
BuRR achieves $\Oh{1}$ query time by choosing $h(k)$ such that its non-zeroes are within a small window.
More specifically $h(k) = 0^{s(k)-1} c(k) 0^{m-s(k)-w+1}$, where $w ∈ ℕ$ is called the \emph{ribbon width} (e.g.\ $w = 64$), $s(k)\in \{1,\dots,m-w+1\}$ is a random starting position, and $c(k) ∈ \{0,1\}^w$ is a random binary string.
Hence, to compute $h(k)Z$, BuRR has to consider at most $w$ consecutive entries of $Z$, beginning with entry $s(k)$.
In addition to improving cache efficiency during queries, this locality (and sorting the rows by $s(k)$) transforms the system of linear equations we have to solve during construction into a form that is almost a diagonal matrix:
the only 1-bits form a narrow \emph{ribbon} along the diagonal, as we illustrate in \cref{fig:burr}.
However, this locality comes at the cost of the solvability of the system $h(k)Z=f(k)$.
If too many keys have similar starting positions $s(k)$, the system becomes unsolvable.
BuRR then \emph{bumps} keys recursively to another BuRR data structure to reduce the number of keys in overloaded regions and therefore ensure solvability.
A fixed number of consecutive starting positions form a \emph{bucket} (see \cref{fig:burr}) and share bumping metadata.
The metadata stores a threshold that splits the keys in the bucket into those that remain in the equation system and those that are bumped.
In the presence of bumping, it is advantageous to choose $m$ slightly \emph{smaller} than $n$. This ensures that every part of the data structure is at full capacity.
See~\cite{DillingerHSW2022burr} for details.

\subsection{VL-BuRR}
\label{ss:comp-retr-vlr}

\begin{figure}
    \centering
    \includegraphics[scale=0.9]{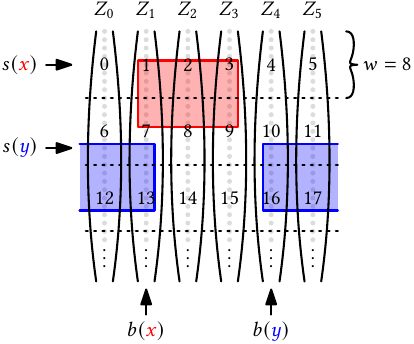}
    \caption{\boldmath
        A VL-BuRR data structure using $\lmax$ separate $1$-bit SFs, given by column vectors $Z₀,…,Z_{\lmax-1}$ (here $\lmax = 6$).
        Shaded areas are two examples of the bits accessed by a single query.
        Grey dots indicate individual bits, dotted lines indicate how the bits are grouped into words of length $w$, and numbers indicate the order in which these words are stored.
    }
    \label{fig:vl-burr}
\end{figure}

In \cref{sec:prelim}, we explained how VL-SFs can be constructed by encoding each bit as an individual key.
An alternative technique is to use several independent 1-bit SFs, one for each bit position of the values, and to query them bit by bit.
However, these straightforward approaches suffer from poor cache locality and thus high query latency.
We address this by adopting an interleaved storage layout of $\lmax$ instances of 1-bit BuRR, where $\lmax$ is the length of the longest value.
There are three ingredients needed to accomplish~this.
\begin{enumerate}
    \item \emph{Same Load.}
    There are a total of $B=\sum_{k\in K} |f(k)|$ bits that our VL-SF has to represent.
    We ensure that all of our $\lmax$ BuRR instances represent roughly equal numbers  of bits (\mbox{$\approx B/\lmax$}) as follows.
    The first bit of each value is represented by the $b(k)$-th SF, where $b: K \rightarrow \{0,1,\dots,\lmax-1\}$ is a hash function.
    The second bit is represented by the $((b(k) + 1) \bmod \lmax)$-th SF and so on.
    We therefore view the individual BuRRs arranged in a cyclic list.
    \item \emph{Same Dimension.}
    To ensure that the solution vectors $Z_i$ of all instances have equal size, we set their sizes to the maximum $m_{\max}$ of what the individual sizes would naturally be.
    \item \emph{Same Hash Function.}
    We choose the same hash function $s(k)$ in each structure to map keys to starting positions.
\end{enumerate}
Let $Z_i\in \{0,1\}^{m_{\max}}$ be the solution vector of the $i$-th BuRR data structure, where $i\in \{0,1,\dots,\lmax-1\}$.
With the ingredients above, the bits that a query reads are in the same positions within all $Z_i$.
To make the queries cache efficient, we only have to interleave the storage layout of all $Z_i$.
We do this by dividing each $Z_i$ into groups of $w$ bits (where $w$ is the ribbon width).
We then store the groups from all $Z_i$ in a round-robin fashion.
Usually, all the bits that a query needs to read are therefore in the same cache line.
We give an illustration in \cref{fig:vl-burr}.

BuRR ensures that the resulting system of linear equations is solvable by bumping keys to a fallback structure.
Note that when used in a variable-length context, bumping information across the BuRR structures is positively correlated: if the $i$-th bucket of the $k$-th ribbon requires bumping, it is quite likely that the $i$-th bucket of the $(k+1)$-th ribbon also requires bumping, because of the codes that extend across both structures.
Given this observation, it is wasteful to store bumping information for each BuRR structure individually.
Instead, the $i$-th buckets of all structures share common bumping information.
Like in $r$-bit BuRR, bumped keys are handled recursively using a second instance of our data structure.

\section{Our LSF based on VL-B\MakeLowercase{u}RR}
\label{s:lsf}

In this section, we present our construction of learned static functions.
Recall the plan to rely on an ML model to output a distribution $μ_k$ on $V$ for each $k ∈ K$ and to use $μ_k$ as a prior when storing and retrieving $f(k)$.
In the following, we describe this auxiliary data structure~$\mathcal{D}$ that stores all $f(k)$.
Roughly speaking, we construct a binary prefix-free code based on $μ_k$ and store the codeword for $f(k)$ in the VL-SF from \cref{s:comp-retr}.

An issue with prefix codes generally is that for $v \sim μ$ the length $ℓ_v$ of its codeword may have an expectation $L=\mathbb{E}[ℓ_v]$ that significantly exceeds the entropy $H(μ)$.
\Cref{fig:suboptimal-prefix-codes} illustrates three such cases.
The first case shows that $H(μ)$ could be close to $0$ while $L$ is always at least $1$.
This can happen if a value has a probability close to~$1$.
Previous work on CSFs has addressed this problem with a filter that stores all keys \emph{not} assigned the most common value~\cite{HreinssonKP09compressedfun,ShibuyaBK22kmer,ColemanRLLS2023caramel}.
Most keys that \emph{do} use the most common value can then be recognised by being negative elements of the filter.
We use this “filter trick” and expand it to cases where a skewed binary decision occurs in places other than the code tree's root node\changed{, which can save up to $\frac{1}{2}$ bits per key} (second case in \cref{fig:suboptimal-prefix-codes}).
\changed{In our experiments, this expansion saved up to \fiterrootspace~bits per key.}
Our techniques actually allow us to work with suboptimal prefix codes (third case in \cref{fig:suboptimal-prefix-codes}) without suffering from worse space efficiency.

\begin{table}
    \caption{\boldmath Code trees (leaves annotated with probabilities) where the expected codeword length $L$ significantly exceeds the entropy $H(μ)$ of the underlying distribution.}
    \label{fig:suboptimal-prefix-codes}
    \begin{tabular}{c@{}ccc}
        \toprule
        \begin{minipage}[b]{1.7cm}\centering
            Code tree\\and\\distribution~$μ$
        \end{minipage}
        & \includegraphics[scale=0.9,page=1]{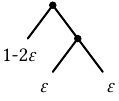} & \includegraphics[scale=0.9,page=2]{fig/prefix-trees} & \includegraphics[scale=0.9,page=3]{fig/prefix-trees}\\
        \midrule
        $L$
        & $1+2ε$ & $\frac 32$ & $2-2ε$\\
        $H(μ)$ & $\Oh{ε\log 1/ε}$ & $1+\Oh{ε \log 1/ε}$ & $1+\Oh{ε \log 1/ε}$\\
        $\lim\limits_{ε → 0}\frac{
        L
        }{H(μ)}$ & $∞$ & $50\%$ & $∞$\\
        \bottomrule
    \end{tabular}
\end{table}

We begin in \cref{ss:weighted-filters} by explaining how VL-BuRR can serve as a \emph{weighted filter}. We then use the weighted filter to build in \cref{ss:WRM} what we call a \emph{weighted relative membership} data structure.
This data structure can efficiently store the skewed binary choices made when descending the code tree. By suitably concatenating the information related to the same key, we obtain our LSF implementation in \cref{ss:LSF-construction}.
Finally, in \cref{ss:comp-retr-code} we discuss our use of suboptimal prefix codes, which are much faster to construct on the fly than optimal prefix codes (e.g.\ Huffman) and significantly improve query time with minimal sacrifices in space efficiency.

\subsection{Weighted Filters from VL-SF}
\label{ss:weighted-filters}

Recall that a filter is constructed for a set $M$ and answers queries of the form ``$k \in M$?'' with no false negatives and a false positive probability of $ε$.
It has long been known \cite{DP:Succinct:2008} that a filter with $ε = 2^{-r}$ can be obtained from an $r$-bit SF as follows.
We use a hash function $g: U → \{0,1\}^r$ (sometimes called \emph{fingerprint function}) and build an SF for the restriction $f = g|_M$ of $g$ to the domain~$M$.
A query simply checks whether SF.query$(x) = g(x)$, which is true for all $x ∈ M$ by construction, and true for $x ∈ U \setminus M$ only with probability $2^{-r}$ since $g(x)$ is a random bitstring independent from~the~SF.

The idea naturally generalises to \emph{weighted filters}. These are constructed for a set $M$ and a function $r : U → ℝ_{≥ 0}$ to which we assume oracle access. A query must satisfy
\[
    \Pr[\text{query}(x) = 1] \begin{cases}
        = 1 & \text{ if $x ∈ M$,}\\
        ≤ 2^{-r(x)} & \text{ if $x ∉ M$.}
    \end{cases}
\]
If $r$ only attains values in $ℕ₀$ then we can implement such a filter by storing a fingerprint of length $r(x)$ in a VL-SF.

Previous work has studied weighted Bloom filters and variations in contexts where the elements of $U$ differ in their probabilities of being in $M$ and differ in their probabilities of being queried \cite{bruck:weightedBloom:2006,bercea:daisyBloom:2024}.

\subsection{Weighted Relative Membership from Weighted Filters}
\label{ss:WRM}

A relative membership data structure (RM) \cite{BelazzouguiV13compressedfun} is constructed for two sets $M,X$ with $m = |M|$, $n = |X|$, and $M ⊆ X ⊆ U$. For a given $x ∈ X$, it must answer whether $x ∈ M$. Equivalently, an RM is a $1$-bit SF with domain $X$ that indicates membership in $M$.

A \emph{weighted relative membership} data structure (WRM) is given, in addition to $M$ and $X$, a function $p : U → [0,\frac 12]$, which we interpret as probabilistic information indicating $p(x) = \Pr[x ∈ M \mid x ∈ X]$.\footnote{The assumption that $p(x) ≤ \frac 12$ simplifies further discussion. It is made without loss of generality as the role of “$∈$” and “$∉$” can be swapped for keys $x$ with $p(x) > \frac 12$.}
For now, let us assume that $M$ is indeed obtained by including each $x ∈ X$ independently with probability $p(x)$.
The information-theoretic space lower bound for the WRM is then $\sum_{x ∈ X} H(p(x))$ bits, where $H(p) = p\log₂(1/p)+(1-p)\log₂(1/(1-p))$ is the binary entropy function.
The following WRM gets close to this bound.

We construct a weighted filter $F$ for $M$ (with weight function $r: U → ℕ₀$ defined below) as well as a $1$-bit SF that stores for all positives of $F$ whether they are true positives or false positives. We give pseudocode in \cref{alg:WRM-constr,alg:WRM-query}.

\begin{algorithm}[t]
\SetAlgoLined
\KwIn{$M,X,p: U \rightarrow [0,\frac 12]$}
\KwOut{a WRM data structure}
choose $r: U → ℕ₀$ \tcp{see discussion}
$F ← \text{constructWF}(M,r)$\;
truePos $← \{x ↦ 1 \mid x ∈ M\}$\;
falsePos $← \{x ↦ 0 \mid x ∈ U \setminus M, F.\text{query}(x) = 1\}$\;
SF $←$ constructSF(truePos $∪$ falsePos)\;
\Return ($F$,SF)\;
\caption{Construction of our WRM data structure.}
\label{alg:WRM-constr}
\end{algorithm}

\begin{algorithm}[t]
\SetAlgoLined
\KwIn{($F$,SF), $x$}
\Return \textbf{if} $F$.query($x$) \textbf{then} SF.query($x$) \textbf{else} $0$\;
\caption{Querying our WRM data structure.}
\label{alg:WRM-query}
\end{algorithm}

We now discuss which choice of $r$ minimises space. We assume for simplicity that the $1$-bit SF requires $1$ bit per key and the weighted filter requires $r(x)$ bits for each $x ∈ M$ (with $1$-bit BuRR and VL-BuRR we get within \vlburrover\% of this ideal on sufficiently large inputs where lower order terms are negligible).

Any $x ∈ X$ contributes $p(x)·r(x)$ bits to the expected space usage of $F$ (it requires $r(x)$ bits if it is stored, which happens with probability $p(x)$) and it contributes $p(x)+(1-p(x))2^{-r(x)}$ bits to the static function (with probability $p(x)$ it is a true positive of $F$, with probability $(1-p(x))2^{-r(x)}$ it is a false positive of $F$). The optimal choice of $r(x)$ depends on $p(x)$. We define
\def\rOptR{r^*_{ℝ_{≥0}}}
\def\rOptN{r^*_{ℕ₀}}
\begin{align*}
    \text{space}(p,r) &= pr+p+(1-p)2^{-r}\\
    \rOptR(p) &= \argmin_{r ≥ 0} \text{space}(p,r)\\
    \rOptN(p) &= \argmin_{r ∈ ℕ₀} \text{space}(p,r)
\end{align*}
When given a (hypothetical) space-optimal weighted filter that supports non-integer $r$, we should choose $r(x) = \rOptR(p(x))$.\footnote{A (non-weighted) filter supporting non-integer $r$ is described in \cite[\S4]{HreinssonKP09compressedfun}. It is not quite space-optimal, though.} 
Given that our BuRR-based weighted filter only supports integer $r$, we choose $r(x) = \rOptN(p(x))$.
In \cref{fig:filterEff}, we plot the resulting space overhead compared to $H(p)$ in both cases. Under the idealising assumptions we made, the space overhead of our approach compared to $H(p)$ is at most $≈ 10.8\%$.

\begin{figure}[t]
    \centering
    \input{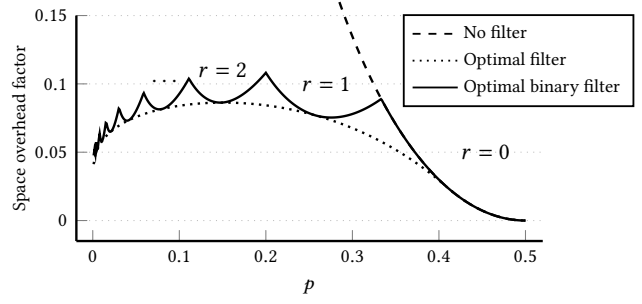}
    \caption{\boldmath Idealised space overhead of filter-based WRM for keys with weight $p ∈ [0,\frac 12]$. \emph{Optimal filter} shows the overhead $(\text{space}(p,\rOptR(p))-H(p))/H(p)$ when using a weighted filter that supports real-valued weights. The maximum is $\approx 0.086$ at $p \approx 0.15$. \emph{Optimal binary filter} shows the overhead $(\text{space}(p,\rOptN(p))-H(p))/H(p)$ when using a weighted filter that supports integer weights only. The maximum is $\approx0.108$ at $p = 0.2$.}
    \label{fig:filterEff}
\end{figure}

\paragraph{Learned Relative Membership and Calibration\@.}
Now assume $p: U → [0,\frac 12]$ arises from an ML model. We obtain a data structure for \emph{learned relative membership} or simply a $1$-bit LSF (to be generalised shortly).

The arguments just given, including the bound of $≈ 10.8\%$ on space overhead, still apply in this case if the underlying model is \emph{calibrated}. Calibration means that out of all keys from $X$ for which the model outputs a probability of roughly $p$, roughly a $p$-fraction are in fact in $M$. The average contribution of many keys to overall space consumption then behaves as predicted by our probabilistic analysis, even though $M$ is not random. Our space goal, previously the entropy $\sum_{x ∈ X} H(p(x))$, is now formally the surprisal $\sum_{x ∈ M} \log₂(1/p(x))+\sum_{x ∈ X \setminus M} \log₂(1/(1-p(x))$ of $M$ given $p$. If $p$ is calibrated, these quantities are (roughly) the same.

When using non-calibrated models, the space overhead of our data structure can be higher than predicted. The problem can be mostly mitigated by choosing $r(x)$ randomly (e.g. rounding $\rOptR(p(x))$ up or down while preserving expectation). But given that our loss function incentivises calibration and since we observe good calibration in our experiments, we have not investigated this issue in detail.

\subsection{LSFs via Weighted Relative Membership: The Generalised Filter Trick}
\label{ss:LSF-construction}

We now describe our construction of LSFs, first focusing on the conceptual level and then on efficient implementation.

\paragraph{Conceptual Level\@.} Recall that for each key $k ∈ K$, we obtain a probability distribution $μ_k = ℳ(k)$ on $V$ from the ML model $ℳ$. For each $k$, we construct a code tree $T = T(k)$ and a corresponding codeword $c₁…c_ℓ ∈ \{0,1\}^*$ with $ℓ = ℓ(k)$ that describes how the leaf with label $f(k)$ is reached in $T$.
Instead of storing $c₁,…,c_ℓ$ in a VL-SF in plain form, we effectively apply the filter trick (explained in \cref{sec:prelim}) at every node of $T$ to save space whenever a binary decision is unbalanced.

Formally, for each node $u$ of $T$, we define $u.p$ to be the combined probability (under $μ_k$) of all leaf labels in the subtree of $u$. If $u$ is an inner node with children $u₀$ and $u₁$, this also gives probabilities $u₀.p/u.p$ of “going left” at $u$ and $u₁.p/u.p$ of “going right” at $u$.
In this sense, the binary decisions $c₁,…,c_ℓ$ leading to $f(k)$ are associated with probabilities $p₁,…,p_ℓ$.
Note that the product $p = p₁·…·p_ℓ$ is simply the probability of $f(k)$ under $μ_k$.

Let $X$ be the set of $\sum_{k ∈ K}ℓ(k)$ binary decisions relating to the keys from $K$. Let $M ⊆ X$ be the subset of those decisions that are made in the less likely way (breaking ties consistently). In principle, we can use the WRM data structure described in \cref{ss:WRM} to store $M$ and recover the prefix code of any $k ∈ K$ using $ℓ(k)$ queries.
The amortised contribution of $k ∈ K$ to overall space is (under the assumptions we have made, including calibration) at most $10.8\%$ more than the sum $\sum_{i = 1}^{ℓ(k)} \log₂(1/p_i)$ of the surprisal of the individual binary decisions. This sum is equal to $\log₂(1/p)$, which is the surprisal of $f(k)$ given $μ_k$. We have hence reached the goal we have set in the introduction (see \cref{eq:loss-function}).

\paragraph{Implementation\@.} A remaining challenge is to implement this idea while interleaving the $ℓ$ requests to the WRM to improve query efficiency.
Recall that a WRM query for $x ∈ X$ first requests some number $r = r(x)$ of filter bits from a weighted filter and, if the filter bits match, another bit from a $1$-bit SF to correct for false positives.
An LSF query for $k ∈ K$ therefore involves a number $F = r₁+…+r_ℓ$ filter bits ($r_i$ bits for the $i$th binary decision) and up to $ℓ$ correction bits.
The correction bits can simply be concatenated and stored using VL-BuRR.
For the filter bits, we modify VL-BuRR to aggregate many weighted filter queries into a single query as follows.

At construction time, we specify for each $k ∈ K$ a bit string $s ∈ \{?,1\}^F$ where $F$ depends on $k$. 
A query for $k ∈ K$ returns a random bitstring $t ∈ \{0,1\}^{F_{\max}}$ (for a fixed $F_{\max} ≥ F$) obtained from $s$ by replacing each $?$ with a random bit and appending $F_{\max}-F$ random bits.
Recall that VL-BuRR distributes the task of storing a bitstring $f(k) ∈ \{0,1\}^F$ to $F$ independent $1$-bit SFs with interleaved storage.
We can distribute $s ∈ \{?,1\}^{F}$ in the same way except that a “$?$” causes the corresponding SF to be skipped and not store anything. To ensure that a random bit is returned from a skipped SF, we use the standard trick of masking the stored bits with random~bits.%
\footnote{We use an additional hash function $g: U → \{0,1\}^{F_{\max}}$. Before construction, we replace $s ∈ \{?,1\}^F$ with $s' ∈ \{?,0,1\}^F$ by replacing each $1$ with the corresponding bit from $g(k)$. When a query would return $t' ∈ \{0,1\}^{F_{\max}}$, it instead returns $t ∈ \{0,1\}^{F_{\max}}$ that indicates bitwise equality of $t'$ with $g(k)$ (with $C$-style operators: \texttt{$t$ = $t'$ \textasciicircum\ $\sim$ $g(k)$}).
}

In \Cref{alg:constr} and \cref{alg:query}, we show how our LSF can be built from these components. We suppress $w_{\max}$ and assume that VL-BuRR returns streams of output bits from which we read as needed.
Note that at query time, the length $ℓ$ of the prefix code of $f(k)$, the numbers $r₁,…,r_ℓ$ of filter bits, and the number $C$ of correction bits are only revealed gradually as we descend the~code~tree.

\def\filterBits{\mathrm{filterBits}}
\def\correctionBits{\mathrm{correctionBits}}
\def\filterStream{\mathrm{filterStream}}
\def\correctionStream{\mathrm{correctionStream}}

\begin{algorithm}[ht]
\SetAlgoLined
\KwIn{$K$, $f: K \rightarrow V$}
\KwOut{LSF}
Train $ℳ$ on $f$ with loss function $S$\;
filterData $← \{\}$ \tcp{empty dictionary}
\For{$k \in K$}{
    $T ←$ code tree for $μ_k = ℳ(k)$\;
    $c₁…c_ℓ ←$ codeword for $f(k)$ in $T$\;
    $(u₀,…,u_ℓ) ←$ path to leaf with label $f(k)$ in $T$\;
    filterBits $← (\,)$\;
    \For{$i = 1$ \KwTo $ℓ$}{\
        \tcp{$u_i.p$ is sum of the probabilities of values in $u_i$'s subtree}
        $p ← u_i.p / u_{i-1}.p$ \tcp{probability for $c_i$}
        $r \gets \text{optimalBitLength}(p)\in \mathbb{N}_0$\;
        \tcp{add to filter if $c_i$ is the less probable value}
        filterBits.append(\textbf{if} $p < 0.5$ \textbf{then} $1^r$ \textbf{else} $?^r$)
    }
    filterData[$k$] $←$ \text{filterBits}\;
}
\,\tcp{data structure aggregating weighted filter queries (see Sec.\,\ref{ss:LSF-construction}):}
filterVL-SF $←$ constructVL-BuRR*(filterData) \tcp{*specialised}
correctionData $← \{\}$ \tcp{empty dictionary}
\For{$k \in K$}{
    $T ←$ code tree for $μ_k = ℳ(k)$\;
    $c₁…c_ℓ ←$ codeword for $f(k)$ in $T$\;
    $(u₀,…,u_ℓ) ←$ path to leaf with label $f(k)$ in $T$\;
    correctionBits $← (\,)$\;
    filterStream $\gets$ filterVL-SF.query($k$)\;
    \For{$i = 1$ \KwTo $ℓ$}{
        $p ← u_i.p / u_{i-1}.p$ \tcp{probability for $c_i$}
        $r \gets \text{optimalBitLength}(p)\in \mathbb{N}_0$\;
        \If{\upshape filterStream.read$(r) = 1^r$}{
            correctionBits.append($c_i$)
        }
    }
    correctionData[$k$] $←$ correctionBits\;
}
correctionVL-SF $←$ constructVL-BuRR(correctionData)\;
\Return $ℳ$, filterVL-SF, correctionVL-SF\;
    \caption{Construction of our LSF.}
\label{alg:constr}
\end{algorithm}

\begin{algorithm}[ht]
\SetAlgoLined
\KwIn{$k\in K$, LSF($ℳ$, filterVL-SF, correctionVL-SF)}
\KwOut{value $\in V$}
$T ← $code tree for $μ_k = ℳ(k)$\;
filterStream $←$ filterVL-SF.query($k$)\;
correctionStream $←$ correctionVL-SF.query($k$)\;
$u ←$ root of $T$\;
\While{$¬u$.isLeaf}{
    $p ←$ $u$.rightChild.$p / u.p$\;
    $r ← \text{optimalBitLength}(p)\in \mathbb{N}_0$\;
    \uIf{\upshape filterStream.read$(r) = 1^r$}{
        $c ←$ correctionStream.read(1)\;
    }\Else{
        $c ←$ (\textbf{if} $p > 0.5$ \textbf{then} $1$ \textbf{else} $0$)\;
    }
    $u ←$ (\textbf{if} $c = 1$ \textbf{then} $u$.rightChild \textbf{else} $u$.leftChild)\;
}
\Return $u$.label\;
    \caption{Query of our LSF.}
\label{alg:query}
\end{algorithm}

\subsection{A Fast Prefix Code}
\label{ss:comp-retr-code}
When querying a key $k$, we construct a prefix code using the probability distribution $\mu_k$ given by the model.
Preliminary experiments show that using optimal Huffman codes is expensive because they have to fully sort the value probabilities \cite{huffman2007method}.

\paragraph{Shannon Codes\@.}\label{par:shannon-codes}
We identified Shannon codes \cite{shannon1948mathematical} as a good compromise between construction time and prefix code lengths.
One reason why Shannon is faster than Huffman is that we only sort the values by their code length $l(p):=\ceil{-\log(p)}$.
The code length is directly determined by the exponent $e(p)$ of the floating point representation of the probabilities using $l(p)=-e(p)$.
Similar to bucket sort, the values are directly assigned to buckets, where each bucket is responsible for one code length.
Starting with the values of shortest code length, we assign the next available codeword of its desired length to each value as follows.
Let $p_1, \dots, p_{|V|}$ be the probability of the values under $\mu_k$, ordered by ascending code lengths.
To compute the codeword for the $i$-th key, we first calculate $\sum_{j=1}^{i-1} 2^{-l(p_j)}$ and then use the leading $l(p_i)$ bits in the fractional part of this value as the code.

We have previously argued that our bound of roughly $10.8\%$ on the space overhead holds regardless of the choice of prefix code. 
In a preliminary experiment using the datasets and models described in \cref{sec:experiments}, we compare Shannon with Huffman codes and find that Shannon codes increase the space by at most \huffshanspace\% \changed{while improving query time by \huffshanspeedup\% on average.
Although our primary objective is space reduction, we include this optimisation because it achieves a substantial speedup at only a marginal space cost.}

\paragraph{Descending an Implicit Code Tree\@.} At query time, we descend the code tree and have to determine the probability of the next bit, i.e. the probability of going left or right (see \cref{alg:query}). All we need is the sorted sequence of values with associated codewords corresponding to the leaves of the code tree. A node $u$ in the code tree is implicitly given by its depth $d$ and two indices \mbox{$1 ≤ i ≤ j ≤ |V|$} indicating a range of leaves. The probability of going left or right at $u$, as well as the index $s ∈ [i,j]$ separating the two subtrees, can be computed with a linear scan of the range $[i,j]$. It is easy to see that this process takes $\Oh{|V|}$~time. %

\paragraph{Avoiding Code Construction\@.}
If one value has a probability $p>50\%$, we can apply an optimisation during queries.
In this case, we can try not to construct a prefix code initially.
Because the probability is more than $50\%$, we immediately know that the code of the most likely value is ``0'' and that the probability that the first bit is ``0'' is simply $p$, and we therefore know how many bits of the variable-length filter code to look at.
Only if we run into the unlikely case where the first bit turns out to be a ``1'', we do have to actually construct the prefix code.
This optimisation is particularly helpful if the model often confidently predicts the correct symbol.
Preliminary experiments show that this trick improves the query time by up to \filteredfanospeedup\% on our datasets.

\section{Experiments}
\label{sec:experiments}
In this section, we evaluate the performance of our LSF on real-world and synthetic datasets.
Our experiments show that:
\begin{itemize}
  \item LSFs can leverage key-value correlations in the data to break the zero-order entropy barrier, achieving a space reduction between \minspacerw \% and \maxspacerw \% (corresponding to \urlspacefactor$\times$ less space) compared to the best existing compressed static functions (CSF) on real-world datasets.

  \item Our auxiliary data structure~$\mathcal{D}$ is highly space efficient, offering a multiplicative space overhead below \storageover \% over the ideal space $S(ℳ,f)$.

  \item Query performance of LSFs remains practical with an average of \allquerygov$\times$ slower queries compared to the CSF competitor.

  \item Even when using our LSF without learning, i.e. when using it as a CSF, it remains more space efficient compared to the CSF competitor, which shows the effectiveness of our VL-BuRR structure and the generalised filter trick.
\end{itemize}

The code to reproduce our experiments is publicly available at \url{\vldbavailabilityurl}.

\subsection{Experimental Setup}

\paragraph{Hardware and Software\@.}
We use a Rocky Linux 9.5 machine with an Intel Core i7-11700 CPU with 64~GiB of DDR4 RAM.
\changed{The CPU is set to a fixed clock frequency of 2.5~GHz.}
Each core has 48~KiB L1 and 512~KiB L2 data cache.
We compile using clang 17 and compiler options \texttt{-march=native} and \texttt{-O3}.

We compare our LSF with the state-of-the-art compressed static function GOV~\cite{GenuzioOV20minimal} and the uncompressed static function BuRR~\cite{DillingerHSW2022burr}.
The GOV and BuRR implementations are from the original authors.
Note that the GOV construction is implemented in Java, but the queries are executed in C\texttt{++}.
For BuRR, we use the configuration recommended for fast queries.
Query time is measured on 10 million randomly chosen keys, and taking the average of 10 runs.%
\footnote{We note that the experimental setting has been revised from the original publication~\cite{HermannLVW26}: GOV now hashes only an integer key, rather than the key and the features required for the learned models, as the other non-learned static functions (BuRR/CSF), thus resulting in directly comparable query times. The revised results and main paper takeaways remain consistent with the original findings.}

\paragraph{Datasets\@.}
We use 4 real-world datasets from various domains, differing in number of input features, output classes, and entropy of the values, along with 4 synthetic datasets.
\Cref{tab:datasets}~summarises~them.
\begin{itemize}
    \item \textit{covertype.}
        The task is to map a land to one of 7 forest cover types (such as pine, spruce, etc.) using 54 cartographic variables (elevation, soil type, distance to water, etc.)~\cite{covertype}.
    \item \textit{nids.}
        The task is to map a network flow to one of 10 cyber attack types (such as DDoS, ransomware, etc.) using 10 features extracted from packet captures (protocol, incoming and outgoing number of bytes or packets, etc.)~\cite{nids}.
    \item \textit{songs.}
        The task is to map a song to one of \songsclasses~song genres using numerical audio features (such as tempo, loudness, instrumentalness, etc.)~\cite{songs}.
    \item \textit{urls.}
        The task is to store 1 bit indicating whether a URL is legitimate or phishing using 15 features extracted from the URL itself (such as use of HTTPS, number of letters, digits, special characters, etc.)~\cite{phiusiil}.
    \item \textit{gauss.}
        The task in this synthetic dataset is to map a scalar value to one of \gaussclasses~classes, where each class corresponds to a Gaussian distribution centred at a location along the real line. The class means are equally spaced with a distance of~2. The standard deviation~$\sigma$ controls the degree of overlap between the distributions, thus the hardness of the task. We vary $\sigma \in \{0.25,0.5,0.75,1\}$. The number of samples is equal across all classes.
\end{itemize}

We keep manual preprocessing to a minimum, applying only one-hot encoding for categorical variables and standard scaling for numerical ones.
We do not modify the real dataset distribution with class balancing techniques commonly used in ML, as they would trivially maximise the entropy and make CSFs ineffective.

\begin{table}[tb]
	\caption{Dataset statistics.}
	\label{tab:datasets}
	\centering
	
\centering
	\begin{tabular}{lrrrr}
		\toprule
		Dataset         & $n$     & \# Features & \# Classes & $H_0$  \\ \midrule

  covertype &  581012 & 54 &  7 & 1.739 \\
       nids & 1379274 & 10 & 10 & 2.376 \\
      songs & 1159764 & 26 & 82 & 6.249 \\
       urls &  235795 & 15 &  2 & 0.985 \\
      gauss &  100\;M &  1 &  8 & 3.000 \\

		\bottomrule

	\end{tabular}

\end{table}

\paragraph{Model Choice\@.}\label{par:model-choice}

For our LSF implementations, we employ lightweight models, namely logistic regression~(LR), Gaussian Naive Bayes~(GNB), and compact multi-layer perceptrons~(MLPs), which, as we will show, offer both time and space efficiency.
For MLP, we consider three configurations by varying the number of hidden layers~$L$ and hidden units~$U$ per layer as $(L,U)\in\{(1,50),(1,100),\allowbreak(2,50)\}$.
We use the softmax activation function for the output layer and ReLU for the hidden layers.
We train the LR and MLP models in Keras using the Adam optimiser and minimise the surprisal $S(ℳ,f)$ as defined in \cref{eq:loss-function}.
Training stops if the surprisal did not improve by at least 1\% within 3 epochs on the validation~set.

The LR and MLP models are then exported to the TensorFlow Lite format and invoked from our LSF implementation in C\texttt{++}.
In addition to experimenting with \texttt{float32} weights, we apply post-training quantisation to compress them to \texttt{float16} and \texttt{int8}.

For GNB, we use our own training and inference implementation in C\texttt{++}.
We apply GNB only to the synthetic gauss dataset, as its generation process aligns directly with the inductive bias of GNB.

For all models, we report both accuracy and top-3 accuracy (for the multi-class datasets) on a hold-out test set comprising 20\% of the data selected via stratified sampling.
From the remaining training data, we use 10\% as a validation set to enable early stopping during the training of LR and MLP models.

In our experiments, the model parameters and the training time are always accounted for in the performance of the LSF. However, in scenarios where a model is already available (e.g., for an upstream task), we note that it could be reused without incurring additional storage cost.
Conversely, we observe that our training setup encourages model generalisation.
This makes the resulting model potentially reusable across multiple LSFs on similar data, thus amortising both model training and storage costs.
Interestingly, in additional experiments with early stopping based on the training loss (i.e., deliberately overfitting the data), we observed almost identical results, most likely because the small models do not have enough capacity to memorise the training data.

\begin{table*}[t]
	\caption{\boldmath
		Model performance, quantised to \texttt{float16}.
        Training \changed{and inference time} are measured in \textmu{}s per key, and space consumption in bits per key.
        For each dataset, models are ordered by total space $\Sigma = S(ℳ,f) + |enc(ℳ)|$.
	}
	\label{t:modelPerformance}
	
\centering

	\begin{tabular}{l ccr rrrrrrr}
		\toprule
        \multirow{2}{*}{Dataset ($H_0$)} & \multicolumn{3}{c}{Model} & \multirow{2}{*}{Training} & \multicolumn{2}{c}{Accuracy} & \multicolumn{3}{c}{Space}& \multirow{2}{*}{Inference} \\
                                           \cmidrule(lr){2-4}                                      \cmidrule(lr){6-7}             \cmidrule(lr){8-10}
                                         & Layers & Units     & \# Param       &                          & Top       & Top-3            & $S(ℳ,f) / n$ & $|enc(ℳ)| / n$ & $\Sigma / n$&  \\

  \midrule\multirow{4}{*}{covertype (1.74)} &                  2 & 50 &  5657 & 215 &  87.4 & 99.9 & 0.4361 &  0.1558 & 0.5919 & 1.489 \\
                                            &                 1 & 100 &  6207 & 220 &  85.2 & 99.9 & 0.5133 &  0.1709 & 0.6842 & 1.345 \\
                                            &                  1 & 50 &  3107 & 168 &  81.8 & 99.8 & 0.6202 &  0.0856 & 0.7057 & 1.078 \\
                                            &  \multicolumn{2}{c}{LR} &   385 &  53 &  72.2 & 99.3 & 0.9172 &  0.0106 & 0.9278 & 0.563 \\
       \midrule\multirow{4}{*}{nids (2.38)} &                  2 & 50 &  3610 &  57 &  70.2 & 94.2 & 1.0578 &  0.0419 & 1.0997 & 1.418 \\
                                            &                  1 & 50 &  1060 &  53 &  69.4 & 93.3 & 1.1020 &  0.0123 & 1.1143 & 1.003 \\
                                            &                 1 & 100 &  2110 &  53 &  69.5 & 93.6 & 1.0931 &  0.0245 & 1.1176 & 1.290 \\
                                            &  \multicolumn{2}{c}{LR} &   110 &  39 &  66.4 & 91.4 & 1.2762 &  0.0013 & 1.2775 & 0.557 \\
      \midrule\multirow{4}{*}{songs (6.25)} &                  2 & 50 &  8082 &  57 &  29.4 & 50.4 & 3.8438 &  0.1115 & 3.9553 & 1.838 \\
                                            &                  1 & 50 &  5532 &  54 &  28.5 & 49.1 & 3.9243 &  0.0763 & 4.0006 & 1.400 \\
                                            &                 1 & 100 & 10982 &  56 &  29.4 & 50.3 & 3.8526 &  0.1515 & 4.0041 & 1.959 \\
                                            &  \multicolumn{2}{c}{LR} &  2214 &  31 &  23.8 & 43.0 & 4.3365 &  0.0305 & 4.3671 & 1.016 \\
       \midrule\multirow{4}{*}{urls (0.99)} &  \multicolumn{2}{c}{LR} &    16 & 217 &  99.4 &   -- & 0.0448 &  0.0011 & 0.0459 & 0.248 \\
                                            &                  1 & 50 &   851 &  94 &  99.7 &   -- & 0.0243 &  0.0577 & 0.0820 & 0.775 \\
                                            &                 1 & 100 &  1701 &  76 &  99.6 &   -- & 0.0244 &  0.1154 & 0.1398 & 1.105 \\
                                            &                  2 & 50 &  3401 &  38 &  99.6 &   -- & 0.0253 &  0.2308 & 0.2561 & 1.194 \\
         \midrule gauss $\sigma$=0.25 (3.0) & \multicolumn{2}{c}{GNB} &    16 &  <1 & 100.0 &   -- & 0.0002 & <0.0001 & 0.0002 & 0.166 \\
                   gauss $\sigma$=0.5 (3.0) & \multicolumn{2}{c}{GNB} &    16 &  <1 &  96.0 &   -- & 0.1526 & <0.0001 & 0.1526 & 0.180 \\
                  gauss $\sigma$=0.75 (3.0) & \multicolumn{2}{c}{GNB} &    16 &  <1 &  84.0 &   -- & 0.5579 & <0.0001 & 0.5579 & 0.185 \\
                     gauss $\sigma$=1 (3.0) & \multicolumn{2}{c}{GNB} &    16 &  <1 &  72.2 &   -- & 0.9139 & <0.0001 & 0.9139 & 0.157 \\

		\bottomrule

	\end{tabular}

\end{table*}

\begin{table}[t]
    \caption{\boldmath
        Comparison with competitors.
        LSF construction time includes training, space usage includes the ML model (quantised to \texttt{float16}), and query time includes model inference.
        From \cref{t:modelPerformance}, we select  models minimising space $\Sigma$ \changed{or  inference time}.
        L and U denote the number of hidden layers and hidden units.
        ``Ours (CSF)'' refers to our LSF without learning, i.e. the model outputs the value frequencies.
        On the gauss dataset, non-learned approaches perform independently of~$\sigma$ and are thus reported once for ``any $\sigma$''.
    }
    \label{t:fullDsPerformance}
    
\addtolength\tabcolsep{-1.9pt}
\centering

	\begin{tabular}{ll rrr}
		\toprule
        \multirow{2}{*}{Dataset ($H_0$)} & \multirow{2}{*}{Competitor} & Constr. & Space    & Query \\
                                         &                             & \textmu{}s/key  & bits/key & \changed{\textmu{}}s/key  \\

  \midrule\multirow{5}{*}{covertype (1.74)} & \textbf{Ours} (L2 U50) & 217.970 & 0.6360 & 1.696 \\
                                            &     \textbf{Ours} (LR) &  54.489 & 1.0050 & 0.816 \\
                                            &    \textbf{Ours} (CSF) &   0.630 & 1.8020 & 0.115 \\
                                            &                    GOV &   1.883 & 1.9403 & 0.063 \\
                                            &                   BuRR &   0.132 & 3.0471 & 0.019 \\
       \midrule\multirow{5}{*}{nids (2.38)} & \textbf{Ours} (L2 U50) &  60.254 & 1.1573 & 1.732 \\
                                            &     \textbf{Ours} (LR) &  40.859 & 1.3564 & 0.861 \\
                                            &    \textbf{Ours} (CSF) &   0.752 & 2.4451 & 0.133 \\
                                            &                    GOV &   1.751 & 2.5642 & 0.059 \\
                                            &                   BuRR &   0.131 & 4.0541 & 0.026 \\
      \midrule\multirow{5}{*}{songs (6.25)} & \textbf{Ours} (L2 U50) &  64.431 & 4.1005 & 3.478 \\
                                            &     \textbf{Ours} (LR) &  37.601 & 4.5231 & 2.693 \\
                                            &    \textbf{Ours} (CSF) &   1.676 & 6.3102 & 0.134 \\
                                            &                    GOV &  13.243 & 6.5181 & 0.062 \\
                                            &                   BuRR &   0.129 & 7.0823 & 0.033 \\
       \midrule\multirow{4}{*}{urls (0.99)} &     \textbf{Ours} (LR) & 218.011 & 0.0526 & 0.391 \\
                                            &    \textbf{Ours} (CSF) &   0.507 & 1.0180 & 0.043 \\
                                            &                    GOV &   1.671 & 1.0941 & 0.042 \\
                                            &                   BuRR &   0.123 & 1.0266 & 0.010 \\
              
\midrule

  \multirow{1}{*}{gauss $\sigma$=0.25 (3.0)} & \textbf{Ours} (GNB) & 0.347 & 0.0006 & 0.300 \\
   \multirow{1}{*}{gauss $\sigma$=0.5 (3.0)} & \textbf{Ours} (GNB) & 0.515 & 0.1665 & 0.361 \\
  \multirow{1}{*}{gauss $\sigma$=0.75 (3.0)} & \textbf{Ours} (GNB) & 0.722 & 0.6066 & 0.415 \\
     \multirow{1}{*}{gauss $\sigma$=1 (3.0)} & \textbf{Ours} (GNB) & 0.800 & 0.9845 & 0.404 \\

\midrule\multirow{3}{*}{gauss any $\sigma$ (3.0)}

   & \textbf{Ours} (CSF) & 1.317 & 3.0159 & 0.105 \\
   &                 GOV & 2.030 & 3.1485 & 0.084 \\
   &                BuRR & 0.129 & 3.0454 & 0.038 \\

		\bottomrule

	\end{tabular}

\end{table}

\subsection{Internal Evaluation}
\label{ss:inteval}

As a first step of our evaluation, we identify the best-performing model of each dataset.
We then integrate the model into our LSF and evaluate it as a whole.

Generally, all models captured the relation between keys and values, as the remaining surprisals $S(ℳ,f)$ are significantly below the zero-order entropy of the values of each dataset.

\paragraph{Weight Quantisation\@.}\label{par:weight-quantisation}
We found that quantising the model weights to \texttt{float16} results in at most just \quantover\% higher surprisal $S(ℳ,f)$ than \texttt{float32} for all datasets and models.
At the same time, the model requires about half the size $|enc(ℳ)|$.
The \texttt{int8} quantisation results in an even smaller model but has a drastically worse performance in terms of overall space $\Sigma := S(ℳ,f) + |enc(ℳ)|$, which is up to \quantoverint\% higher.
\changed{Regarding inference time, \texttt{float16} quantisation is \quantinfover\% faster than \texttt{float32} on average; instead, \texttt{int8} is \quantinfoverint\% slower than \texttt{float32}, likely because our models are shallow and any potential speedup is offset by type conversion overhead.}
We therefore use the \texttt{float16} quantisation henceforth.

\paragraph{Model Performance\@.}
\Cref{t:modelPerformance} shows the training time, accuracy, space consumption, and inference time of the \texttt{float16} models in detail.
On the real-world covertype, nids, and songs datasets, the MLP model with 2 hidden layers and 50 hidden units performed best in overall space consumption $\Sigma$. 
Interestingly, the songs dataset is the most challenging in terms of learnability, as evident from its lower model accuracies and higher surprisal, and yet $\Sigma$ is up to \songsentropysigma\% lower than $H_0$, which highlights the effectiveness of LSFs even in scenarios where the model predictions are less confident.
In contrast, the real-world urls dataset is highly learnable: all models predict the correct values with high confidence, as evidenced by their high accuracy and the low values of $S(ℳ,f)$. 
In such cases, the model size $|enc(ℳ)|$ could become the dominant term in the total space $\Sigma$.
Accordingly, the simple LR model performed best here, despite having a value $S(ℳ,f)$ nearly twice as high as that of more complex models.
On all datasets, we observe differences in training and inference times across models.
Hence, selecting a model entails a trade-off between space and time that directly affects the performance of the overall LSF.
\changed{In the subsequent evaluation of the overall LSF (\Cref{ss:comparison-competitors,t:fullDsPerformance}), we select for each dataset both the model that minimises $\Sigma$ and the one that minimises the inference time, thereby illustrating both ends of the trade-off.}

\paragraph{Model Calibration\@.} In \cref{ss:WRM}, we discussed that the space efficiency of WRM (underlying our LSFs) relies on a well-calibrated model.
A model is well-calibrated if events of predicted probability $p$ actually occur a $p$ fraction of the time.
A common metric to measure the calibration of a model is the expected calibration error (ECE) \cite{degroot1981assessing, naeini2015obtaining}.
An ECE of 0\% indicates a perfectly calibrated model, while an ECE of 100\% indicates full miscalibration.
We measured an ECE of at most \maxeceperc \% for the model of each dataset (using 50 equally-sized bins, see \cite{naeini2015obtaining} for details). %

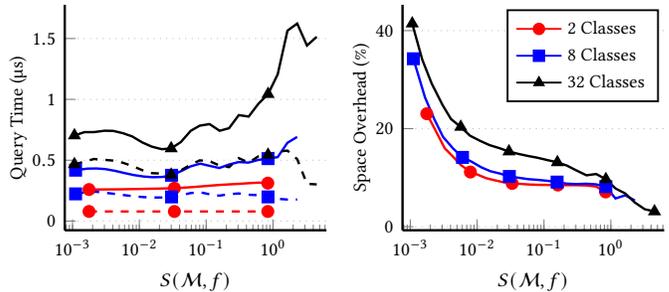
\begin{figure}[t]
	\begin{subfigure}[h]{0.22\textwidth}
		    \begin{tikzpicture}
        \begin{axis}[
            title={},
            xlabel={$S(ℳ,f)$},
            height=3cm,
            ylabel={Query Time (\textmu{}s)},
            every axis plot post/.append style={mark repeat=10},
            xmode=log,
            cycle list={
                {colorClasses2, solid, mark=*},
                {colorClasses2, dashed, mark=*},
                {colorClasses8, solid, mark=square*},
                {colorClasses8, dashed, mark=square*},
                {colorClasses32, solid, mark=triangle*},
                {colorClasses32, dashed, mark=triangle*},
            },
          ]

          \addplot coordinates { (0.001766,0.207558) (0.002494,0.207038) (0.003277,0.207693) (0.004252,0.207735) (0.005722,0.208539) (0.007926,0.209167) (0.010308,0.210174) (0.013754,0.210894) (0.018701,0.211466) (0.02487,0.213268) (0.033446,0.214687) (0.045206,0.21629) (0.061058,0.219558) (0.083902,0.223902) (0.115795,0.227529) (0.161395,0.232751) (0.223101,0.237301) (0.308217,0.241785) (0.428408,0.245743) (0.598406,0.24768) (0.835387,0.240698) };
          
          \addplot coordinates { (0.001766,0.0578154) (0.002494,0.0577887) (0.003277,0.0577937) (0.004252,0.0577961) (0.005722,0.0578268) (0.007926,0.057826) (0.010308,0.0578393) (0.013754,0.0577924) (0.018701,0.0577899) (0.02487,0.0577851) (0.033446,0.0578047) (0.045206,0.0578078) (0.061058,0.0577898) (0.083902,0.0577904) (0.115795,0.0577882) (0.161395,0.0577994) (0.223101,0.0577927) (0.308217,0.0577891) (0.428408,0.0577944) (0.598406,0.0578005) (0.835387,0.0577913) };
          
          \addplot coordinates { (0.001112,0.372584) (0.001661,0.384568) (0.002281,0.386338) (0.003088,0.382471) (0.00436,0.366627) (0.00604,0.345984) (0.00822,0.328064) (0.01153,0.314014) (0.015828,0.307579) (0.022184,0.307386) (0.030344,0.319435) (0.042015,0.350605) (0.058494,0.388847) (0.081367,0.405078) (0.112447,0.38202) (0.157127,0.360976) (0.219178,0.38984) (0.306826,0.408176) (0.42935,0.397871) (0.600352,0.412162) (0.841511,0.41209) (1.173789,0.405988) (1.642692,0.494566) (2.29725,0.537624) };
          
          \addplot coordinates { (0.001112,0.202537) (0.001661,0.215791) (0.002281,0.216611) (0.003088,0.208818) (0.00436,0.197273) (0.00604,0.18681) (0.00822,0.17728) (0.01153,0.169683) (0.015828,0.165806) (0.022184,0.165835) (0.030344,0.170718) (0.042015,0.184412) (0.058494,0.203252) (0.081367,0.209104) (0.112447,0.191836) (0.157127,0.180073) (0.219178,0.193066) (0.306826,0.197666) (0.42935,0.18542) (0.600352,0.179692) (0.841511,0.161931) (1.173789,0.137959) (1.642692,0.136871) (2.29725,0.136926) };
          
          \addplot coordinates { (0.001069,0.584961) (0.001506,0.611851) (0.002088,0.609755) (0.00303,0.621778) (0.004067,0.621618) (0.005688,0.605722) (0.007811,0.578676) (0.011064,0.531255) (0.015329,0.49769) (0.021695,0.476613) (0.029942,0.482115) (0.041847,0.531678) (0.058814,0.621022) (0.080188,0.656767) (0.112188,0.66865) (0.158765,0.614925) (0.219799,0.632181) (0.307345,0.728253) (0.432276,0.714776) (0.604497,0.80768) (0.842856,0.86792) (1.175717,0.994724) (1.644532,1.27296) (2.300793,1.30868) (3.219987,1.09432) (4.502753,1.11923) };

          \addplot coordinates { (0.001069,0.380391) (0.001506,0.401758) (0.002088,0.403132) (0.00303,0.415122) (0.004067,0.409164) (0.005688,0.395156) (0.007811,0.369411) (0.011064,0.336076) (0.015329,0.312265) (0.021695,0.297434) (0.029942,0.299666) (0.041847,0.327437) (0.058814,0.382607) (0.080188,0.413347) (0.112188,0.409966) (0.158765,0.356605) (0.219799,0.365301) (0.307345,0.433644) (0.432276,0.401091) (0.604497,0.452365) (0.842856,0.464766) (1.175717,0.490898) (1.644532,0.497185) (2.300793,0.429052) (3.219987,0.23573) (4.502753,0.231654) };
        \end{axis}
    \end{tikzpicture}
	\end{subfigure}
    \hfill
	\begin{subfigure}[h]{0.22\textwidth}
		    \begin{tikzpicture}
        \begin{axis}[
            title={},
            xlabel={$S(ℳ,f)$},
            height=3cm,
            ylabel={Space Overhead (\%)},
            every axis plot post/.append style={mark repeat=5},
            cycle list={
              {colorClasses2,solid,mark=*},
              {colorClasses8,solid,mark=square*},
              {colorClasses32,solid,mark=triangle*},
            },
            xmode=log,
          ]

          \addplot coordinates { (0.001766,23.0464) (0.002494,19.1259) (0.003277,15.9902) (0.004252,14.4403) (0.005722,12.583) (0.007926,11.1532) (0.010308,10.5937) (0.013754,9.88076) (0.018701,9.44335) (0.02487,9.07117) (0.033446,8.83215) (0.045206,8.72672) (0.061058,8.60493) (0.083902,8.54092) (0.115795,8.51418) (0.161395,8.50212) (0.223101,8.47284) (0.308217,8.47844) (0.428408,8.41488) (0.598406,8.20262) (0.835387,7.09456) };

          \addlegendentry{2 Classes};
          
          \addplot coordinates { (0.001112,34.2626) (0.001661,26.3697) (0.002281,21.9641) (0.003088,18.329) (0.00436,15.9633) (0.00604,14.1391) (0.00822,13.2847) (0.01153,12.1162) (0.015828,11.3154) (0.022184,10.8276) (0.030344,10.2491) (0.042015,9.97739) (0.058494,9.70014) (0.081367,9.55547) (0.112447,9.36352) (0.157127,9.10537) (0.219178,8.7819) (0.306826,8.69483) (0.42935,8.77186) (0.600352,8.64326) (0.841511,8.181) (1.173789,5.76194) (1.642692,6.40686) (2.29725,5.34526) };

          \addlegendentry{8 Classes};
          
          \addplot coordinates { (0.001069,41.4406) (0.001506,34.1301) (0.002088,29.2146) (0.00303,24.8515) (0.004067,21.9572) (0.005688,20.3586) (0.007811,18.5764) (0.011064,17.5163) (0.015329,16.6482) (0.021695,16.0129) (0.029942,15.3397) (0.041847,14.8398) (0.058814,14.4098) (0.080188,14.0694) (0.112188,13.5968) (0.158765,13.1339) (0.219799,12.4778) (0.307345,11.4653) (0.432276,10.4887) (0.604497,10.7516) (0.842856,9.67947) (1.175717,7.804) (1.644532,6.7725) (2.300793,4.91252) (3.219987,3.56259) (4.502753,3.10919) };

          \addlegendentry{32 Classes};
        \end{axis}
    \end{tikzpicture}
	\end{subfigure}
	\caption{\changed{\boldmath Query time, inference time (dashed) and space overhead on the gauss dataset varying \#~classes and $S(ℳ,f)$.}}
	\label{fig:internalf}
\end{figure}

\paragraph{Structure Overheads\@.} Besides the model $ℳ$, another part of our LSF is an auxiliary data structure~$\mathcal{D}$ that intuitively makes up for what $ℳ$ does not know about $f:K\rightarrow V$ (see \cref{s:lsf}).
Ideally, this internal data structure would require about $S(ℳ,f)$ bits of space.
However, as \cref{t:fullDsPerformance} shows, it requires slightly more space due to two main factors:
(i)~the WRM data structure underlying our LSF has an overhead of up to 10.8\% (as determined in \cref{ss:WRM}, assuming $ℳ$ is calibrated), and
(ii)~VL-BuRR has additional overhead, e.g. for storing bumping data, amounting to at most \vlburrover\% on large inputs.
To measure the overhead introduced by the internal data structure, we compare its space with the ideal space $S(ℳ,f)$.
On most datasets, the overhead is below \storageover \%, indicating that the worst-case overhead of 10.8\% rarely occurs in practice.
Exceptions to the overhead of \storageover\% occur for the gauss $\sigma=0.25$ and the urls dataset, in which the overhead is \storageovergausszero \% and \storageoverurls \%, respectively. 
In both cases, the LSFs are so effective that their overall size is less than \spacemaxgausszerourls\;KB, making modest constant overheads of the data structure (such as internal pointers and counters) appear large, while in fact remaining negligible in absolute terms \changed{(see also \cref{fig:internalf})}.

\paragraph{Construction\@.}
The construction time strongly depends on the training complexity of the model.
For simple models like GNB, training times are below \gausstrainmax~\changed{\textmu{}}s per key, which is negligible in the overall construction of the LSFs.
Complex models like MLP (L2 U50) on the covertype dataset result in training times as high as \covertrainmus~\textmu{}s per key and dominate the overall LSF construction.
Training time can be significantly reduced through standard ML techniques like smaller training sets, increased batch sizes, early stopping, etc., and is of secondary concern for our algorithmic evaluation of LSFs. 

\paragraph{Queries\@.}
Model inference time makes up between \infminr{}\% and \infmaxr{}\% of the total query time \changed{on the most space efficient models and at most \infmaxrfast{}\% on the fastest models}.
Overall, the query time of our LSF ranges from \querymin{}~to \querymax{}~\changed{\textmu{}}s per key across all datasets (see \cref{t:fullDsPerformance}).
The query time positively correlates with the number of classes and $S(ℳ,f) / n$ \changed{(see \cref{t:fullDsPerformance} and \cref{fig:internalf})}.
This is due to the increased inference time required to compute probabilities for more classes and the higher cost of constructing the prefix code in each query.
Finally, the traversal time of the code tree positively correlates with the average~code~length which is roughly $S(ℳ,f) / n$.

\subsection{Comparison With Competitors}\label{ss:comparison-competitors}
In \cref{t:fullDsPerformance}, we compare the performance of our LSF with the CSF GOV~\cite{GenuzioOV20minimal} and the $r$-bit SF BuRR~\cite{DillingerHSW2022burr}, which are both introduced in \cref{sec:prelim}.
The table also includes a new CSF based on our LSF, which we compare against GOV at the end of this section.

Qualitatively, we find that our LSF significantly outperforms GOV and BuRR in terms of space consumption.
This is not surprising, because the space of our LSF undercuts the zero-order empirical entropy $H_0$ of the value sequence on all datasets, while the space of GOV is lower bounded by $H_0$.
The $r$-bit SF BuRR has an even higher lower bound of $r=\lceil \log(|V|) \rceil$ bits per key.
Looking at specific datasets, our LSF is particularly effective on the urls and the synthetic gauss datasets.
Our LSF achieves a space reduction by at least \urlspace\% compared to BuRR and GOV on the urls dataset.
Using the most space efficient model, space reduction is \covertypespace\% and \nidsspace\% compared to the closest competitor on the covertype and nids dataset, respectively.
The songs dataset has the largest number of classes (\songsclasses) and is the hardest to learn in the sense that $S(ℳ,f)$ is the closest to the value entropy.
Still, our LSF results in \songsspaceperc\% less space compared to the closest competitor.

On the synthetic gauss dataset, we varied the variance $\sigma$ of the \gaussclasses~classes.
Since the class means are fixed at a distance of 2, lower variances $\sigma$ allow the model to be more confident about the value of a key.
For $\sigma=1$, our LSF requires \gausthreespace\% less space compared to the closest competitor.
Pushing $\sigma$ down to $0.25$, we achieve a space reduction of \gauszerospace\% compared to the closest competitor, i.e. the closest competitor requires \gauszerospacefactor$\times$ as much space as our LSF.

\paragraph{Construction\@.}
In terms of construction time, our LSF is at most \consallgov~times slower compared to GOV and \consallburr~times slower than $r$-bit BuRR across all datasets.
Training by far dominates the construction time of our LSF.
However, we remark that on larger datasets or in tasks that can reuse an existing model (e.g. upstream tasks), training times might not be an issue.
Some scenarios might also allow for simpler models.
One example of this is the gauss dataset, where we use GNB.
GNB has a fast training time because it only has to determine the mean and variance of each value.
On all gauss datasets, we are even faster in construction than GOV with a speedup of at least \consgaussgov$\times$.
This is also because of the high compression that we can achieve, resulting in only a small amount of data that the LSF has to encode.

\paragraph{Queries\@.}
In terms of queries, our LSF has to perform model inference and constructs a prefix code for each query.
GOV has no model and constructs the prefix code only once.
BuRR even uses fixed-length codes.
The slower query speed of our LSF is therefore expected.
Over all datasets, query times of the fastest LSF variants are on average \allquerygov$\times$ slower than GOV and \allqueryburr$\times$ slower than BuRR.

\paragraph{Our CSF\@.}
We replaced the machine learning model of our LSF with one that outputs the value frequencies, independently of the key.
Our LSF then becomes a CSF, allowing us to evaluate the performance of our generalised filter trick and our VL-BuRR data structure without the learning part.
Instead of constructing per-key prefix codes on demand, we build a global Huffman code shared by all keys, improving query performance.
The space of our data structure, like any CSF, is now lower bounded by the value entropy.
The resulting performance is reported in \cref{t:fullDsPerformance} under the name ``Ours~(CSF)''.
On all datasets, our CSF achieves space usage within \ourcsfover\% of the value entropy, with the VL-BuRR component contributing no more than \ourcsfburrover\% to this overhead.
The remaining space overhead is due to sub-optimal coding, which we further reduce compared to GOV using our generalised filter trick.
Our CSF outperforms the state-of-the-art GOV competitor in terms of construction time and space on all datasets.
We achieve a space reduction as high as \ourcsfgov\% compared to GOV on the covertype dataset.

Note that our CSF outperforms the BuRR competitor in space on the urls and gauss datasets, where $H_0\approx \ceil{\log_2(|V|)}$, even though we also use BuRR internally.
This is because we use a space-efficient BuRR configuration internally, while the competitor uses a faster configuration, as speed is its main selling point in our comparison.

\section{Related Work}\label{sec:related}
Several learning-based approaches have been proposed for data structure design and lossless data compression.
Here, we mention approaches that are conceptually related to LSF.

DeepMapping~\cite{DeepMapping} memorises a key-value map with a neural network and stores the misclassified pairs in key-sorted partitions, which are then compressed with a standard general-purpose compressor.
Hence, unlike an LSF, it explicitly stores the keys and does not exploit the model's output probabilities for compression.

Sequential neural  compressors~\cite{SchmidhuberH96,DeletangRDCGMGW24,DeepZip} compress sequences such as texts by combining arithmetic coding with a neural network (e.g. a language model) that predicts a probability distribution for the next symbol given the preceding context.
As such, they do not provide a randomly-accessible key-value map.

Some integer compressors approximate the input sequence with a regression model and encode the residuals, either supporting random access~\cite{Ao:2011,Boffa2022talg,FerraginaMV2022linearities,Guerra2025neats,leco2024} or not~\cite{Stearns95,StearnsTM93}.
Unlike LSFs, they neither use a probabilistic classifier nor provide a key-value map.

Learned Bloom filters (LBFs)~\cite{KraskaBCDP2018learned,Mitzenmacher18,RaeBL19,AdaBF,VaidyaKMK21,ReviriegoHDS21,DaiSRH22,SatoM23,PatgiriBN23,MalchiodiRFGF24} train a binary classifier to accept keys from a given set $K \subseteq U$ and reject keys from $U \setminus K$, and use a backup Bloom filter to eliminate false negatives, i.e. keys from $K$ that the classifier incorrectly rejects.
LSFs are more general than LBFs in that they map keys to arbitrary values rather than just a binary membership.
Even in the binary case, LSFs differ from LBF in that they guarantee no false positives on a given set of non-keys from $U \setminus K$. 
Learned Functional Bloom filters (LFBF)~\cite{ByunL22} extend LBFs to return a value associated with each key, rather than only approximate membership information. However, they introduce indeterminables, i.e. queries for which the data structure cannot determine a value at all.
Thus, LFBFs offer different guarantees than LSFs.

Perfect hash functions \cite{Knuth1998art,lehmann2025modern} (PHFs) are related to SFs as they also do not need to store the keys.
A PHF for a set~$K$ of size~$n$ maps the keys from~$K$ to a range~$\{1,…,m\}$ for some~$m ≥ n$ without collisions.
An SF can be derived directly from a PHF~\cite{DBLP:journals/jacm/FredmanKS84,DBLP:conf/stacs/HagerupT01}.
Conversely, SFs have been used to construct PHFs \cite{BPZ:Practical:2013,LehmannSW2022sichash,DBLP:conf/alenex/Lehmann0W24}.
Monotone Minimal PHFs (MMPHFs) specialise PHFs by mapping each key in $K$ to its rank in $\{1,…,n\}$~\cite{BelazzouguiBPV2011theoryPractice,assadi2023tight,FerraginaLSV2023lemon}. MMPHFs have also been studied in the learned setting~\cite{FerraginaLSV2023lemon}, where a learned index~\cite{FerraginaV2020pgm} is used as a rank estimator.
A different line of work replaces traditional hash functions with learned models in standard hash tables~\cite{sabek2022can,KraskaBCDP2018learned}.
However, hash-based indexing techniques do not address the compression of values associated with keys.

\section{Conclusion}
\label{sec:conclusion}

\pgfmathsetmacro{\gaussfactor}{int(100/(100-\gauszerospace))}
\pgfmathsetmacro{\urlfactor}{int(100/(100-\urlspace))}

We have introduced the concept of a \emph{learned static function} (LSF). Like a compressed static function (CSF), it is meant to store key-value assignments in contexts where some values are more likely than others. \emph{Unlike} CSFs, however, the prediction can be based on the requested key itself rather than merely on the global value frequencies in the data set. While the space of CSFs cannot beat the zero-order entropy of the values, our LSFs can be much smaller if there is a learnable correlation between keys and values.

\emph{In principle}, there is no limit on how much more space efficient an LSF can be compared to a CSF (for the synthetic Gauss dataset, the LSF is smaller by a factor of $≈$\gaussfactor).
We show experimentally that this promise is borne out \emph{in practice}: in a real-world dataset with annotated URLs, we save a factor of $≈$\urlfactor.

We pay for this reduced space with increased query times, mostly due to the inference time of the used machine learning model. Whether or not this is acceptable depends on the application context, as does the question of how much the problem can be mitigated, say by relying on simpler heuristics with smaller inference times.

Technically, our LSF construction relies on a combination of machine learning, fast per-key prefix codes, efficient variable-length static functions (VL-SF), and the clever use of weighted Bloom filters. The VL-SF in particular may be of independent interest.

We have suggested potential applications of LSFs in databases, natural language processing, endgame tablebases and bioinformatics. We are curious to see which of these application areas bear fruit and which further directions we may not have anticipated.

\begin{acks}
  This work was supported by funding from the pilot program Core Informatics at KIT (KiKIT) of the Helmholtz Association (HGF).
  GV was supported by the NextGenerationEU -- National Recovery and Resilience Plan (Piano Nazionale di Ripresa e Resilienza, PNRR) -- Project: ``SoBigData.it - Strengthening the Italian RI for Social Mining and Big Data Analytics'' -- Prot. IR0000013 -- Avviso n. 3264 del 28/12/2021.
  We thank Matthias Becht for working on the implementation of VL-BuRR.
\end{acks}

\bibliographystyle{ACM-Reference-Format}
\bibliography{bibliography}

\end{document}